\pdfoutput=1
\documentclass[11pt,preprint]{aastex}
\usepackage{amsmath}
\usepackage{amsfonts}
\usepackage{amssymb}

\def\simlt{\lower.5ex\hbox{$\; \buildrel < \over \sim \;$}}
\def\simgt{\lower.5ex\hbox{$\; \buildrel > \over \sim \;$}}
\def\kms{km s$^{-1}$}

\def\lsun{{L_\odot}}

\def\schi{{\sc Hi}\ }
\def\muH{\mu_{\rm H}}

\def\fir{F_{\rm IR}}
\def\fx{F_{\rm X}}
\def\lir{L_{\rm IR}}

\def\irxobs{$R_{\rm IRX, obs}$}
\def\irxcoll{$R_{\rm IRX, coll}$}
\def\rfifty{$r_{\rm 50}$}

\def\tauion{$\tau_{\rm ion}$}

\def\iras{{\it IRAS}}
\def\spitzer{{\it Spitzer}}
\def\herschel{{\it Herschel}}
\def\chandra{{\it Chandra}}
\def\akari{{\it AKARI}\ }
\def\sst{{\it Spitzer}}
\def\rosat{{\it ROSAT}}
\def\xmm{{\it XMM}}
\def\wise{{\em Wide-Field Infrared Survey Explorer}}
\def\effir{\epsilon_{\rm IR}}
\def\nefv{n_e \sqrt{f_V}}
 
\begin{document}

\shortauthors{Koo et al.}
\shorttitle{Infrared Supernova Remnants}

\title{Infrared Supernova Remnants and their  
Infrared to X-ray Flux Ratios}

\author{Bon-Chul Koo\altaffilmark{1}, Jae-Joon Lee\altaffilmark{2}, 
Il-Gyo Jeong\altaffilmark{1,2}, 
Ji Yeon Seok\altaffilmark{3}, Hyun-Jeong Kim\altaffilmark{1}
}
\altaffiltext{1}{Department of Physics and Astronomy, Seoul National University, Seoul 151-742, Korea; koo@astro.snu.ac.kr}
\altaffiltext{2}{Korea Astronomy and Space Science Institute 776, Daedeokdae-ro, Yuseong-gu, Daejeon 305-348, Korea}
\altaffiltext{3}{Department of Physics and Astronomy, University of Missouri, Columbia, 
MO 65211, USA}
%\altaffiltext{3}{Institute for Astronomy and Astrophysics, Academia Sinica,
%P. O. Box 23-141, Taipei 10617, Taiwan}

\begin{abstract}

Recent high-resolution infrared space missions have revealed 
supernova remnants (SNRs) 
of diverse morphology in infrared (IR) dust emission that is  
often very different from their X-ray appearance. 
The observed range of infrared-to-X-ray (IRX) flux ratios of SNRs are also 
wide. For a sample of 20 Galactic SNRs, 
we obtain their IR and X-ray properties and investigate the physical
causes for such large differences.  
We find that the observed IRX flux ratios (\irxobs) are 
related to the IRX morphology, with SNRs with the 
largest \irxobs\ showing anticorrelated IRX morphology. 
By analyzing the relation of \irxobs\ to X-ray and IR parameters,  
we show that the \irxobs\ of some SNRs agree with theoretical ratios 
of SNR shocks in which dust grains are heated and destroyed 
by collisions with plasma particles.
For the majority of SNRs, however, \irxobs\ values are  
either significantly smaller or significantly larger 
than the theoretical ratios.
The latter SNRs have relatively low dust temperatures.
We discuss how the natural and/or environmental properties of SNRs 
could have affected the IRX flux ratios and the IRX morphology of 
these SNRs. 
We conclude  
that the SNRs with largest  \irxobs\ are probably 
located in dense environment and 
that their IR emission is 
from dust heated by shock radiation rather than by collisions.
Our result suggests that the IRX flux ratio, 
together with dust temperature, can be used to infer the nature of 
unresolved SNRs in external galaxies. 

\end{abstract}


\section{Introduction}

Supernova remnants (SNRs) are one of prominent objects in IR emission. 
\footnote{By ``infrared (IR)'' in this paper, we would generally mean 
``far-infrared'' of wavelength $\simgt 25$  $\mu$m.}
In the Milky Way, among the three hundred known SNRs,
about 20\%--30\% are usually seen in unbiased surveys using the 
the {\em Infrared Astronomical Satellite} (\iras), 
{\em Spitzer Space Telescope}, and \akari\ 
\citep{arendt1989, saken1992, goncalves2011, jeong2012}.
This low fraction, however, is likely mainly 
because the SNRs are located in the Galactic plane
where the confusion from 
foreground and background emission is severe. 
In nearby external galaxies such as 
the Large Magellanic Cloud (LMC), the fraction is much higher, e.g., $\sim60$\% 
\citep{seok2013}. 

A primary mechanism for the IR emission in SNRs 
is thermal dust emission.  
Dust grains swept-up by SN blast waves are heated by collisions with hot X-ray 
emitting plasma and emit mid- and far-infrared emission. 
It has been pointed out that 
the resulting IR dust emission is 
the dominant cooling mechanism for SNRs and 
could be more luminous than the X-ray emission in SNRs over most 
of its lifetime even with dust destruction taken into account 
\citep{ostriker1973, dwek1981, draine1981}.
\cite{dwek1987b} and \citet{graham1987} examined the ratio of 
IR-to-X-ray (IRX) fluxes of 
nine Galactic and four Magellanic SNRs using \iras\ data, respectively, 
and they found that all the examined SNRs have  
ratios greater than unity, so dust IR cooling 
is indeed more efficient than X-ray cooling in SNRs.
However, there was a large scatter in the observed ratios over three orders of 
magnitudes, with a majority of 
SNRs having ratios significantly less than the 
theoretical flux ratio expected for hot gas in thermal equilibrium 
with the dust-to-gas ratio (DGR) 
of the general 
interstellar medium (ISM). \citet{dwek1987b} pointed out that 
the SNRs might have low IRX flux ratios because of 
the destruction of dust grains by SNR shocks or 
the low DGR in the ambient medium, 
while they could have high IRX flux ratios because of the    
extra contribution from IR line emission and emission from 
dust heated by shock radiation.
For large SNRs such as the Cygnus Loop, 
it was shown that the IR emission can be decomposed into two components:
one associated with X-rays, which might be from collisionally heated dust grains, 
and the other associated with optical filaments, 
which could be heated by both collision and radiation within the filament
\citep{arendt1992}.
However, with the \iras\ data, which have a resolution of several arcminutes, 
it was difficult to confirm the true nature of the IR emission for most 
SNRs.

Practical use of the IRX flux ratio for studying dust heating and 
processing has been regained with sensitive, high-resolution 
far-infrared (FIR) data from recent high-resolution IR space missions.
\citet{dwek2008} found that in SN 1987A, the IRX flux ratio decreased significantly 
over three years and showed that the change is consistent with 
the destruction and heating of swept-up circumstellar dust behind the SN shock 
wave. \citet{arendt2010} and \citet{williams2011} derived IRX flux ratios for  
the SNRs Puppis A and RCW 86, respectively, which were found to be 
considerably smaller than the theoretical IRX flux ratio even 
with dust destruction taken into account, and they concluded that  
the DGR 
in the ambient medium of these
SNRs should be lower than that of the general ISM by a factor of several.
It is interesting that, in other studies of Galactic and Magellanic SNRs, 
in which the DGR is obtained by  
directly comparing the dust mass derived from the IR 
SED (spectral energy density) modeling and 
the X-ray emitting mass derived from X-ray spectral analysis
\citep{blair2007,lee2009,borkowski2006,williams2006,williams2011b},
a similar conclusion was reached. \citet{goncalves2011}  
performed a systematic study of the SNRs in the inner Galaxy 
($10^\circ \leq \ell \leq 65^\circ$ and $285^\circ \leq \ell \leq 350^\circ$,  
$|b|<1^\circ$)
using the \sst\ MIPSGAL survey 
(survey of the inner Galactic plane using MIPS; \citet{carey2009})
in which they derived IRX flux ratios of 13 SNRs among 
the 39 SNRs that they detected. The derived IRX flux ratio 
ranged from 1.6 to 240, which confirmed the 
earlier conclusion that the dust IR cooling 
is dominant over the X-ray cooling 
but the result was not discussed further. 
More recently, \cite{seok2015}  
explored the IR and X-ray characteristics of the LMC SNRs, and 
showed that the IRX flux ratios of the 
LMC SNRs are systematically lower than those of the Galactic SNRs
to be consistent with the low DGR in the LMC. 

In this paper, we use the IR and X-ray images from recent space missions,
mainly from the \sst\ and the \chandra\ X-ray telescope, 
to explore the nature of the IR emissions from SNRs and 
the characteristics of their IRX flux ratios. 
These arcsecond-resolution images 
make it possible to pursue the detailed investigation of 
the association between the IR and X-ray emissions. 
We limit our study to Galactic SNRs in this paper. 
The twenty Galactic SNRs studied in this paper 
are composed of eleven\footnote{Among the SNRs studied by \citet{goncalves2011} (see their Table 6),
G33.6+0.1 (Kes 79) and G337.2$-$0.7 are not included 
because these remnants are not clearly visible in the FIR band.}
SNRs studied by \citet{goncalves2011}
and nine other SNRs that are well-defined in FIR band 
(e.g., from 20 $\mu$m to $\simgt 70$ $\mu$m). 
In \S~2, we summarize 
IR and X-ray parameters of these SNRs, introducing an index for the 
IRX morphological correlation. It is found that the IRX flux ratio 
has a systematic dependence on the IRX morphology.  
In \S~3, we compare the observed IRX flux ratios to the theoretical ratios 
expected for hot shocked plasma where dust grains are heated by collisions, 
and we show that time-dependent dust destruction and non-equilibrium ionization 
can explain the observed ratios of some SNRs but not for most SNRs.
This result is discussed in \S~4 together 
with their IRX morphological correlation.
We will see that we can infer the natural and/or environmental characteristics of SNRs from their IRX flux ratios without a prior knowledge on them, which will be very useful to study SNRs in external galaxies where we do not have the luxury of observing their morphology.

\section{Infrared and X-ray Characteristics of Supernova Remnants}

\subsection{Parameters of Infrared Emission}

The parameters characterizing global properties of the 
IR emission from SNRs are the IR flux ($\fir$) and 
color temperature of dust emission ($T_d$, hereafter `dust temperature').
We have derived these parameters for 13 SNRs in this work, including 
11 SNRs studied by \citet{goncalves2011} and
two large SNRs IC 443 and Puppis A. 
For the rest, we adopted the parameters obtained in previous studies.

Eleven SNRs from \citet{goncalves2011} are all covered
by the MIPSGAL survey \citep{carey2009}
and the \herschel\ infrared Galactic Plane Survey (Hi-GAL; Molinari et al. 2010).
The former survey is a legacy program of the {\it Spitzer Space Telescope}
covering the inner Galactic plane
($\ell=0^\circ$--63$^\circ$ and $298^\circ$--$360^\circ$)
in two wavebands (24 and 70 $\mu$m) using MIPS, while
the latter survey is an Open Time Key Project of the {\it Herschel Space Observatory}
covering a similar Galactic longitude range, i.e.,
$\ell=0^\circ$--60$^\circ$ and $300^\circ$--$360^\circ$,
in five wavebands between 70 and 500 $\mu$m using 
the Photodetector Array Camera and Spectrometer (PACS) and the Spectral and Photometric Imaging Receiver (SPIRE).
The latitude coverage is $|b|<1^\circ$
except for $|\ell|<5^\circ$ where the coverage in the MIPSGAL survey is $|b|<3^\circ$.
We used the fully calibrated \spitzer\
MIPS 24 $\mu$m images (Version 3.0) and \herschel\
PACS 70 $\mu$m images (level 2.5), 
which are available from the {\em Spitzer} Heritage 
Archive\footnote{\url{http://sha.ipac.caltech.edu/applications/Spitzer/SHA/}}
and from the {\it Herschel} Science
Archive\footnote{\url{http://www.cosmos.esa.int/web/herschel/science-archive/}},
respectively.
The MIPS 24 $\mu$m and PACS 70 $\mu$m images have comparable spatial resolutions,
i.e., $6''$ and 5.$''2$.
\citet{goncalves2011} derived total IR fluxes of their sources 
using \spitzer\ 24 and 70 $\mu$m fluxes, but we have not used \spitzer\
70 $\mu$m images because they have considerably lower angular resolution ($16''$)
than the \herschel\ 70 $\mu$m images and are often contaminated by artifacts.
The measured 24 and 70 $\mu$m fluxes of the SNRs are listed in Table 1.
For most SNRs, we simply measured the background-subtracted total flux density inside
a circular area surrounding the entire SNR with the background level
estimated from an annulus just outside the circle.
For some SNRs, however,
the contamination from the foreground and/or background emission is dominant,
particularly at 70 $\mu$m, and an accurate total flux density could not be obtained from
such simple analysis.
If, for example, we can measure the total 24 $\mu$m flux density but
can measure the 70 $\mu$m flux density of only a limited region, 
we scaled the measured 70 $\mu$m flux density using
the (70 $\mu$m)/(24 $\mu$m) flux ratios of that area to obtain total 70 $\mu$m flux.
For SNRs that required a special remedy, we have added notes in Table 1.

Two large SNRs, IC 443 and Puppis A, 
have been observed by \cite{noriega2008} and  \cite{arendt2010}, respectively, 
using the \spitzer\ MIPS in fast scan mode. 
We have downloaded the archive images from the \sst\ Heritage Archive
and derived their 24 and 70 $\mu$m fluxes as above.

Total IR fluxes and dust temperatures 
are derived by assuming a single-temperature dust component 
with a power-law opacity of index 2, i.e., 
%\begin{equation}
%F_\nu = \kappa_nu B_\nu(T_d)  
%\end{equation}
%
\begin{equation}
{F_{24} \over F_{70}} = \left(\lambda_{24}\over \lambda_{70} \right)^\beta 
{B_{24}(T_d) \over B_{70}(T_d)}, 
\label{eq-1}
\end{equation}
and
\begin{equation}
F_{\rm IR} = F_{70} {\int \kappa_\nu B_\nu (T_d) d\nu \over \kappa_\nu B_\nu(T_d)}
\label{eq-2}
\end{equation}
where $\kappa_\nu\ (\propto \lambda^\beta$, with $\beta=-2$) is the mass 
absorption coefficient, $\lambda_{24}$ and 
$\lambda_{70}$ are reference wavelengths of the 
24 and 70 $\mu$m fluxes, and $B_{24}(T_d)$ and $B_{70}(T_d)$ are 
the Planck functions at the temperature of $T_d$ at wavelengths $\lambda_{24}$ and 
$\lambda_{70}$, respectively. 
The reference wavelengths for the \spitzer\ 24 and 70 $\mu$m fluxes are 
23.68 and 71.42 $\mu$m, respectively, while it is 70.0 $\mu$m for the \herschel\ 
70 $\mu$m flux. Note that $F_{24}$ and $F_{70}$ in the above equations 
are the {\it color-corrected} fluxes, which differ from the observed fluxes 
usually by no more than a few percent. 
The derived $T_d$ and $F_{\rm IR}$ values are 
listed in Table 1. 
For the SNRs whose parameters are adopted from previous studies, 
the references are given in the last column of the table.
The fluxes in Table 1 should be close to the total IR fluxes of SNRs 
except for RCW 86, for 
which the value is the flux of a small portion of the SNR (see note in Table 1). 

There is one caveat. At 24 and 70 $\mu$m bands,
there are ionic lines which can make a significant contribution to the
observed fluxes, e.g., [Fe II] 24.5 $\mu$m, [Fe II]/[O IV] 26.0 $\mu$m,
[S I] 25.2 $\mu$m, and [O I] 63 $\mu$m lines.
This can happen for SNRs where the shock is radiative.
For Kes 17, for example, the estimated line contributions are
$38\pm10$\% and $16\pm8$\% to the MIPS 24 and
\akari\ 65 $\mu$m fluxes, respectively \citep{lee2011}.
If we derive the IR parameters of this remnant using the continuum fluxes only,
$\fir$ will be $\sim 20$\% lower, and $T_d$ will be 1.5 K lower
than those in Table 1.
\cite{andersen2011} obtained \spitzer\ MIPS spectra of
14 Galactic SNRs with radiative shocks in dense environment,
and Kes 17 has the highest line contribution to the total IR flux.
Therefore, we consider that $\fir$ and $T_d$ in Table 1 could be              
overestimated for some SNRs because of FIR ionic lines but   
the difference is probably not signicant to affect our analysis in this paper.

\subsection{Parameters of X-ray Emission}

Essential X-ray parameters that we need in this study are 
the temperature of the X-ray emitting plasma ($T_e$) and 
soft-X-ray (0.3--2.1 keV) flux ($\fx$). 
Another useful parameter is the ionization timescale
$\tau_{\rm ion}\equiv \int n_e dt$ ($=n_e t$ for a constant electron density $n_e$), 
which is the `time' for the plasma to have reached the current 
ionization state. Both the dust and gas cooling rates, and therefore the 
ratio of IRX fluxes, are  
time-dependent, and \tauion\ can be used as a time parameter 
(see the next section).

X-ray parameters of the 20 SNRs are given in Table 2. 
For SNRs whose parameters had been derived from detailed modeling 
in previous studies, we adopted those parameters. 
The flux values in the literature are sometimes given 
in a different energy range than the range of 0.3 $-$ 2.1 keV adopted in this work. 
For such cases, we converted the flux value in the literature to the 
flux of 0.3 $-$ 2.1 keV by applying a correction factor based on 
the model and the model fit parameters in the literature. 
There are remnants where no suitable flux value was available. 
For these remnants, we extracted spectra of an entire remnant and 
conducted a spectral fitting adopting the 
hydrogen column density of the X-ray absorbing gas ($N({\rm H})$) 
from the literature. 
We note that the unabsorbed flux is insensitive to detailed 
fit parameters except for the $N({\rm H})$ value. 
For three remnants (G11.2$-$0.3, Kes 73 and RCW 103), 
we derived improved fit parameters by conducting a full spectral analysis.
For each remnant, the spectra from smaller regions are extracted and 
are fit with a single non-equilibrium ionization (NEI) model. $N({\rm H})$, $T_e$ and 
\tauion\  in Table 2 represent ensembles of these values. 
To derive the flux, the spectrum of the entire region is extracted 
and fitted with $N({\rm H})$ fixed at the value from the previous step.

\subsection{Infrared and X-ray Morphology of SNRs}

Figure 1 shows the IR 
and X-ray images of the SNRs listed in Tables 1 and 2. 
We see that some SNRs such as 
Kepler, G11.2$-$0.3, and Puppis A have similar  
IR and X-ray morphology,
whereas some SNRs such as 3C391, W49B, and IC 443 have either uncorrelated or  
anticorrelated IR and X-ray morphologies.
To represent their formal resemblance in two wavebands, we use  
Pearson's linear correlation coefficient between the IR and X-ray intensities
 as an index. 
The Pearson's $r$ is defined as the covariance of the 
variables divided by the product of their standard deviations,
and it varies from +1 (perfect correlation) to $-$1 
(perfect anticorrelation). 
It can be obtained straightforwardly
once we define the pixels for correlation, 
e.g., using the IDL procedure CORRELATE.
We use the bright pixels contributing 50\% of total IR or total X-ray fluxes. 
That is, we first examine the cumulative brightness distribution of pixels in 
IR image of a source and determine the brightness level at which the cumulative 
flux becomes 50\% of the total IR flux of the source ($I_{IR, 50\%}$). 
The same can be done for the X-ray image to obtain $I_{{\rm X-ray}, 50\%}$. 
Then the pixels for the correlation will be the ones with {\em either} its 
IR brightness greater than $I_{IR, 50\%}$ {\em or} its X-ray 
brightness greater than $I_{{\rm X-ray}, 50\%}$.
The adopted percentage (i.e., 50\%) is somewhat arbitrary, 
but it yields correlation coefficients 
that reasonably characterize the morphological relation
between the bright IR and X-ray emissions dominating the fluxes.
Before deriving the correlation coefficient, 
we remove strong point sources in the IR and X-ray images and subtract 
a constant background.
We also convolved the IR and X-ray images to the same resolution 
of 6$''$--10$''$ for sources from \chandra/\xmm\ images and 30$''$ for
sources from \rosat\ X-ray images and resampled them onto the same grid
to have the same pixel scale of 1$''$--5$''$ per pixel depending on
the image size.
The computed correlation coefficients \rfifty\
are listed in Table 3, 
while Figure 2 shows the scattered diagrams of IR versus 
X-ray brightnesses used for the computation of \rfifty\ 
for Puppis A and W49B which have 
the largest positive (0.76) and 
the largest negative ($-0.59$) \rfifty, respectively.

The SNRs with relatively large ($\simgt 0.3$) 
positive \rfifty\ are 
Kepler, G11.2$-$0.3, G15.9+0.2, Kes 73, Kes 75, 
the Cygnus Loop, Cas A, Puppis A, MSH 11$-$54,
RCW 86 and RCW 103. 
All these SNRs appear 
to have similar IR and X-ray morphologies in Figure 1.
The SNRs with relatively large ($\simlt -0.3$) 
negative \rfifty\ are 3C391, W49B, W44, and Kes 17.
All these SNRs appear to have anticorrelated 
IR and X-ray morphology in Figure 1.
The SNRs 3C396, 
3C397, Tycho, IC 443, and G349.7 
have \rfifty\ close to 0 (i.e., $-0.16$ to 0.20). 
These SNRs seem to have both correlated and 
anticorrelated components (see Fig. 1). 
We emphasize that 
the correlation coefficients in Table 3 are based on the images in Figure 1 which are 
mostly \spitzer\ 24 $\mu$m and \chandra\ 0.3--2.1 keV images 
(see the caption of Figure 1). 
If there are more than one heating mechanism for dust, 
they could depend on wavelength.

We note that the SNRs with large negative \rfifty\ belong to the category of 
``thermal composite" or ``mixed-morphology" SNRs (hereafter MM SNRs) 
that appear shell or composite type in radio but with 
thermal X-rays inside the SNR \citep{rho1995,rho1998}, e.g., 3C391, W44, and IC 443.
The IR morphologies of these SNRs are very similar to their 
radio morphologies, which explains their large negative \rfifty. 
The correlation coefficient \rfifty\ of the 
MM SNRs, which gives, in a sense, the degree of mixed morphology, 
ranges from $-0.59$ to $-0.15$ excluding G292.0+1.8 which has an 
unusually large positive \rfifty ($=0.46$).
Most of these SNRs are thought to be middle-aged SNRs interacting with dense ISM, 
e.g., molecular clouds, and the origin of center-filling 
thermal X-rays is considered to be   
either dense clumps evaporating inside hot plasma or 
the result of large-scale 
conduction \citep{white1991,cox1999,vink2012}. 
However, not all SNRs interacting with MCs are MM SNRs
and vice versa, so their nature is still 
unclear. For example, G292.0+1.8 is a young SNR of mixed morphology 
because of thermal X-rays from SN ejecta and 
the swept-up circumstellar medium. 
(It also has a large positive \rfifty.)
We will discuss the nature of the SNRs with large negative \rfifty\ 
further in \S~4.

\section{IRX Flux Ratio and Collisional Dust Heating}

In this section, we first obtain the IRX flux ratios of SNRs
and investigate the relation to their IRX morphology. 
We then compare the observed IRX flux ratios to the theoretical ratios expected for hot shocked plasma where dust grains are heated by collisions. We note that the theoretical IRX flux ratios of hot dusty plasma are not applicable for some SNRs, e.g., young SNRs such as Tycho where the X-ray emission is mostly from shocked SN ejecta while the FIR emission is from shocked ambient medium or evolved SNRs such as IC 443 where the FIR emission is mostly from radiatively heated dust. We will see that the IRX flux ratios of such SNRs are significantly different from theoretical IRX flux ratios, so that, even without 
a prior knowledge, one can infer the natural or/and environmental characteristics of an SNR from its IRX flux ratio.

\subsection {IRX Flux Ratio \irxobs}

We define the observed IRX flux ratio as 
\begin{equation}
R_{\rm IRX, obs} \equiv  {F_{IR}\over F_X},
\label{eq-3}
\end{equation}
and the \irxobs\ of 20 SNRs obtained from  
$F_{IR}$ and $F_X$ in Tables 1 and 2 are given in Table 3.
\irxobs\ ranges from 0.32 to 204. 
For comparison, \citet{dwek1987} and \citet{goncalves2011} obtained 
$<2.5$ to 990 and 1.6--240, respectively.
For some SNRs, our values differ from those of previous results
by more than an order of magnitude, e.g., 
ours : those of \cite{goncalves2011} are  
0.32:6.2 (3C397), 15.5:1.6 (RCW 103), and 62.6:5.1 (W49B).
The difference from \cite{goncalves2011} 
values originates mostly from the adopted X-ray fluxes.
\cite{goncalves2011} used the X-ray fluxes in the 
\chandra\ SNR catalog\footnote{\url{
http://hea-www.cfa.harvard.edu/ChandraSNR/}}, some of which need to be 
updated. 
The difference from the values of \cite{dwek1987}, however, 
is partly due to different definitions of IRX flux ratios.
\cite{dwek1987} used X-ray fluxes corrected for NEI effects 
and, for young SNRs such as 
Tycho and Cas A, they used X-ray fluxes from the swept-up ambient medium  
while we simply used total fluxes. 
These factors, which need to be considered in interpreting the IRX flux ratios, 
are discussed in the following sections.
 
%;ours : those of \cite{dwek1987b} = 0.74:60 (Tycho), 
%;1.7:350 (Cas A), 
%13.8:360 (IC 443), and 1.2:990 (RCW86); 
%We also have SNRs with the \irx\ ratio less 
%than 1, i.e., 3C397 (0.32) and Tycho (0.74). 

Figure 3 plots \irxobs\ versus \rfifty. 
We note that the \irxobs\ values of the 
SNRs with negative $r_{50}$ are generally larger than 
those of the SNRs with positive $r_{50}$, or,  
alternatively, the SNRs with high ($\simgt 30$) 
\irxobs\ values have $r_{50}\le 0$.
This suggests that the large scatter could at least 
be partly explained by  
the IR emission from these SNRs being of 
different origin, 
e.g., the SNRs of large negative $r_{50}$ 
and high \irxobs\ values are 
MM SNRs (marked by squares in the figure) 
where the IR emission is 
probably from radiatively-heated dust 
{\em not} from collisionally-heated dust 
in X-ray emitting plasma (see \S~4.2).
Previous studies showed that the \irxobs\ values of SNRs are 
$> 1$, from which they concluded that dust IR cooling dominates 
over the 
X-ray cooling \citep[e.g.,][]{dwek1987, goncalves2011}. 
All except two in our sample have \irxobs$>1$ 
which is consistent with the conclusion of previous studies.  
We have two SNRs, i.e., 3C397 (0.32) and Tycho (0.74), 
that have \irxobs$<1$, but they might have 
additional X-ray emission (see \S~4.1).  

\subsection{Collisional Heating of Dust Grains}

In Figure 4-(a), we plot \irxobs\  
versus the temperature of hot gas $T_e$ for 20 SNRs. 
The figure is similar to Figure 1 of \citet{dwek1987}, but 
there are now 20 SNRs (compared to their 9) and 
their locations in the diagram have been largely shifted.
We first note that there is a large scatter in the observed IRX 
flux ratio:  
SNRs of the same X-ray temperature ($\sim 10^7$~K) 
have \irxobs\ values that differ by more than two orders of magnitude. 
The dotted line in Figure 4-(a)
represents the expected IRX ratios for hot, dusty plasma 
{\em in thermal and collisional equilibrium} \cite[see][]{dwek1987}. 
If the plasma is in thermal equilibrium, its cooling rate 
by dust emission is equal to the heating rate of dust by collisions with 
electrons, so that the dust cooling function 
$\Lambda_d$ (erg cm$^3$ s$^{-1}$) becomes independent of $n_e$ and $n_{\rm H}$ but  
depends on gas temperature $T_e$ and dust composition and size. 
If the plasma is in collisional equilibrium (CIE), 
the X-ray (0.3--2.1 keV) gas cooling function $\Lambda_X$ is also 
a function of $T_e$ only. 
We have calculated the theoretical IRX ratio 
$R_{\rm IRX,coll}\equiv \Lambda_d/\Lambda_X$,  
and the dotted line is for the ISM dust model of \cite{weingartner2001},
who assumed that 70\% of C atoms are locked in graphite grains and all Si atoms are
locked in silicate (MgFeSiO$_4$) grains. 
For the grain size distribution, we adopted 
the MRN size distribution \citep{mathis1977}, 
with minimum and maximum grain radii of $0.001~\mu$m and $0.5~\mu$m, 
respectively. 
For $\Lambda_X$, we have used  the CHIANTI code \citep[V7.1.4; ][]{landi2012}
assuming the solar abundance of \citet{asplund2009}.
We note that, by assuming solar abundance for the hot gas, 
we repeatedly include heavy elements in gas and dust cooling rates. 
However, if \irxcoll\ is to be compared to 
the ratio of IR and X-ray fluxes  
obtained from the entire SNR, this is not an entirely unacceptable 
assumption. Furthermore, the physical condition 
for the shocked gas is expected to be very different from this 
equilibrium condition anyway (see below).
Figure 4-(a) shows that all SNRs except W44 are below the dotted line.

The equilibrium curve in Figure 4-(a), however, is not applicable to 
the IRX flux ratio of SNRs for at least two reasons: 
First, dust grains are destroyed behind the shock front and 
therefore the DGR decreases with time.
Second, it takes time for the swept-up ions to reach CIE, so  
the X-ray cooling rate at the beginning could be  
much lower or higher than that in  the CIE model. 
We have developed a simple plane-shock model 
to calculate time-dependent \irxcoll\ curve.
(Similar models can be found in previous works, e.g., 
see \cite{borkowski2006, williams2006,dwek2008}.)
We assume that dust grains with characteristics 
of the above equilibrium model 
are continuously injected into shocked gas. 
After being injected, they are destroyed by 
thermal and non-thermal sputtering with ions and the 
grain size distribution is modified.
(We neglect grain-grain shattering
which is important for slow radiative shocks.)
The erosion rates ($da/dt$) by sputtering 
depends on the chemical composition and temperature of 
the shocked gas
as well as the speed of grains through the plasma
\citep[e.g.,][]{nozawa2006}. 
We adopt  
 $da/dt= -1.6\times 10^{-6} n_{\rm H}$ and 
 $-4.0\times 10^{-6} n_{\rm H}$ ${\rm \mu m\ yr^{-1}}$ 
 for graphite and silicate grains, respectively. 
These constant erosion rates are 
good within about 50\%  
for the temperature range of our sample SNRs,
i.e., between $2\times 10^6$~K and $2\times 10^7$~K (Table 2).
The gas density 
and temperature behind the shock front are assumed to remain the same. 
We calculated the cooling function of a fluid element 
and obtained a volume-averaged cooling rate by integrating it.
For the X-ray cooling, it is necessary to incorporate 
the increase of the heavy element abundance of shocked gas  
with time owing to dust destruction and 
the different \tauion\ that they experience, 
but we have found that the flux obtained by 
assuming a single \tauion\ gives a satisfactory result.
Figures 5-(a) and (b) show 
the resulting dust and X-ray cooling functions, respectively. 
The dust cooling function starts to deviate from the non-destruction case
at $\tau_{\rm ion}\simgt 10^{11}$~cm$^{-3}$ s, while 
the X-ray cooling function is much lower or higher than the CIE curve 
initially but approaches to it 
at $\tau_{\rm ion}\simgt 10^{12}$~cm$^{-3}$ s.
The resulting time-dependent \irxcoll\ curves are shown as 
solid lines in Figure 4-(b). \irxcoll\ fluctuates 
about the equilibrium ratio initially, but it steadily 
decreases as a result of dust destruction.
As \tauion\ increases from $10^{10}$ to $10^{13}$~cm$^{-3}$ s, 
\irxcoll\ decreases in general at $T_e\simlt 10^7$~K. 
Therefore, the scatter of the observed IRX flux ratio 
in Figure 4-(b) appears to be attributable to an evolutionary effect. 

The \tauion\ values of SNRs derived from the X-ray spectral analysis, however, 
are mostly $10^{11}$~cm$^{-3}$ s except for the MM SNRs (Table 2) 
whereas, if we exclude the SNRs with high negative \rfifty, 
most SNRs excluding the young SNRs 
appear to lie between the two curves 
of $\tau_{\rm ion}=10^{11}$ and $10^{13}$ cm$^{-3}$ s in Figure 4-(b). 
This inconsistency can be also seen in Figure 6 where we plot \irxobs\ versus \tauion\ of SNRs.
If we exclude the SNRs with high negative \rfifty, 
which are well above the theoretical curves,  
the SNRs are all below the theoretical curve and 
there appears to be no obvious correlation between 
\irxobs\ and \tauion.
Hence, the scatter in Figure 4 may be partly due to 
an evolutionary effect, but not entirely. 
We have derived \irxcoll\ values for the individual SNRs from their 
\tauion\ and $T_e$ in Table 2, and the results are listed in   
Table 3. Figure \ref{fig7} shows that the SNRs with negative \rfifty\ 
have $R_{\rm IRX, obs} > R_{\rm IRX, coll}$, whereas most other SNRs 
have $R_{\rm IRX, obs} \simlt  R_{\rm IRX, coll}$. 
In conclusion, only for a small number of SNRs 
are the \irxobs\ values consistent with 
that of hot plasma where dust grains are collisionally heated.

\section{Discussion}

In this section, we first consider possible physical explanations for the 
SNRs with \irxobs\ lower or higher than \irxcoll\ 
(hereafter refered to as SNRs with low and high IRX flux ratios). 
We then discuss the implication on recent results from the LMC SNRs. 

\subsection{Supernova Remnants with Low IRX Flux Ratios}

Among the SNRs with low \irxobs\ values, the most prominent are 
Kepler, Cas A, Tycho, and 3C397. These remnants deviate from the 
\irxcoll\ curve by an order of magnitude in Figure \ref{fig7}. 
The first three are young historical SNRs, and their 
small \irxobs\ values might be due to additional X-ray emission.
In Cas A and Tycho, the X-ray emission is dominated by 
the X-ray emission from SN ejecta and not by 
thermal emission from shocked surrounding gas 
\citep{hwang2012,hwang1997}. In Kepler, 
the X-ray emission from the circumstellar medium is not negligible 
\citep{burkey2013}. We estimate that 
$\ge 50$\% of the X-ray flux 
is from SN ejecta from its soft band (0.3--0.7keV) image.  
It is interesting that these SNRs do have high 
positive \rfifty\ because the IR and X-ray emissions 
originate from spatially distinct
regions unless the SNR contains a large amount of 
newly-formed SN dust such as Cas A. 
However, such positional shifts between the 
IR and X-ray sources are small and not apparent in their \rfifty.   
The nature of another prominent SNR 3C397 is 
uncertain. It had been proposed to be a middle-aged ($\sim 5,300$ yr) 
MM SNR evolving in a molecular cloud \citep{safiharb2005}. 
CO observation revealed a large molecular cloud along the sight line 
toward 3C397 \citep{jiang2010}, although 
there is no direct evidence (e.g., H$_2$ emission) for the interaction. 
However, its hard X-ray emission is 
dominated by SN ejecta, and it has been proposed that 
3C397 is a young Type Ia SNR evolving in a medium of 
low ($\sim 1.6$~cm$^{-3}$) density \citep{yamaguchi2015}.
The location of 3C397 in Figure 7 is 
very different from those of the other evolved MM SNRs 
that have high IRX 
flux ratios, and our result seems to support the 
latter interpretation, i.e., a young type Ia SNR where 
the X-ray emission inside the SNR is largely from SN ejecta. 
However, the X-ray absorbing column is large 
($3\times 10^{22}$~cm$^{-2}$) and there is a nearby HII region northwest 
of the remnant \citep[3C397W;][]{dyer1999} which is bright in infrared (see Fig. 1), so  
this conclusion needs to be confirmed by further studies. 
%we cannot completely rule out that  

There are two other MM SNRs with small \irxobs\ values: G292.0+1.8 and 3C396. 
G292.0+1.8 has significant X-ray emission from SN ejecta \citep{park2004},
which explains its small \irxobs\ value. Note that it has 
\rfifty($=0.47$) very different from other MM SNRs.
For 3C396, we have not found a 
plausible explanation. Its X-ray emission is partly non-thermal,  
but its contribution to the soft X-ray flux appears to be negligible 
\citep{harrus1999}.

%3C396 non-thermal? There could be also some additional X-ray emission from 
%shock-acceleratd high-energy particles and 
%But the contribution from these components should be small
%in soft X-rays.

The other SNRs with considerably small \irxobs\ values are G15.9+0.2, 
Kes 75, Puppius A, and RCW 86. 
These SNRs have large \rfifty\ and thus the FIR emission
is thought to be mainly from collisionally-heated dust.  
Their \irxobs\ values, however, are smaller than those of 
\irxcoll\ by a factor of 2--5. 
(Note that we have excluded the X-ray emission 
from the central pulsar and the pulsar wind nebula in Kes 75.) 
A possible explanation for this difference is that the 
DGR values in the surrounding media of these 
SNRs are lower than that of the general ISM
\citep{arendt2010, williams2011}.
Indeed, a similar conclusion has been reached  
for other Galactic and Magellanic SNRs 
in previous studies
\citep{blair2007,lee2009,borkowski2006,williams2006,sankrit2010,williams2011b}.
Among the SNRs with large \rfifty, G11.2$-$0.3, Kes 73, 
the Cygnus Loop, and RCW 103  
have \irxobs\ consistent with \irxcoll, which appears to indicate that 
the DGR in the ambient or circumstellar medium of these SNRs 
might be equal to that of the general ISM.
It is, however, noteworthy that 
there could be some extra contribution to the IR emission, e.g., 
IR line and/or radiatively heated dust emission (see the next section). 
In the Cygnus Loop, for example, 
\cite{braun1986} estimated that about 25\% of the IRAS 60 $\mu$m 
emission is from the dust in recombined gas in the postshock layer. 
From their analysis of a non-radiative shock in the Cygnus Loop, 
\citet{sankrit2010} concluded that the DGR in the ambient medium is 
about half that of the general ISM. Therefore, it is not 
impossible that these SNRs do also have DGR values somewhat below that 
of the general ISM. 

\subsection{Supernova Remnants with High IRX Flux Ratios}

All the SNRs with \irxobs\ values significantly larger than 
those of \irxcoll\ are 
MM SNRs with large negative \rfifty, e.g., 
W44, W49B, 3C391, IC 443, Kes 17, and G349.7+0.2. 
The \irxobs\ values of some of these SNRs are greater than 
those of \irxcoll\ 
by almost by an order of magnitude. 
The FIR emission in these SNRs 
is not correlated with X-rays, and therefore 
is probably not from collisionally heated dust grains. 
There could be some contribution from ionic lines to the 
observed IR flux,
but, as we discussed in \S~2.1, it is considered to be less 
than a few tens of percent.
Another distinct property of these SNRs indicating a different 
origin of the FIR emission is their low dust temperatures (Figure 8; see also Table 1).
These SNRs have $T_d=43$--51 K whereas 
most other SNRs have $T_d=57$--67 K 
except for the young SNRs Kepler and Cas A. 
For collisionally heated dust in hot ($\simgt 10^7$~K) plasma, 
dust temperature is mainly determined by electron density ($n_e$), and, 
for a typical electron density of $n_e=1$ cm$^{-3}$ and  
silicate grains of radius 0.01--0.1 $\mu$m,  
$T_d=60$--50 K \citep{dwek2008}. 
Therefore, dust temperatures of 57--67 K are consistent with the 
FIR emission being largely from collisionally-heated dust grains of 
radius $\simlt 0.01$~$\mu$m, but 43--51 K might be rather low. 
In Figure 8, there appears to be a weak correlation 
(correlation coefficient $=0.6$) between $R_{\rm IRX,obs}$ and $T_d$,  
and if we naively perform a least-squares fitting excluding 
the SNRs with low \irxobs\ values in Figure 7, we obtain
\begin{equation} 
\log_{10} R_{\rm IRX,obs}=-0.078(0.007)T_d+5.50(0.40),
\label{eq-4}
\end{equation}
where the dust temperature ($T_d$) is in K. 
If dust grains are heated by collisions in hot plasma, then we do not expect such 
a correlation because $T_d$ is mainly determined by $n_e$ whereas 
the IRX flux ratio is independent of $n_e$. Instead, the apparent correlation 
in Figure 8 might represent different origins for the dust IR emissions at 
low and `high' temperatures.

The low dust temperature and large \irxobs\ values of the MM SNRs 
with large negative \rfifty\ in Figure 8
probably indicate that the FIR emission in these SNRs is largely from  
dust heated by shock radiation. 
All these SNRs,
e.g., W44 \citep{reach2005}, W49B \citep{keohane2007}, 
IC 443 \citep{lee2012}, Kes 17 \citep{lee2011}, 
and G349.7+0.2 \citep{lazendic2010},
are indeed known to be interacting with dense ambient medium and 
the shock is largely radiative. 
In a radiative shock, essentially all the shock energy is converted into 
UV and X-ray radiation. This UV and X-ray radiation from the hot shocked gas 
ionizes and heats the surrounding gas to 
produce an extended temperature plateau region in the postshock cooling 
layer and also a `radiative precursor' in the preshock gas, both of which are 
at $T_e\sim 10^4$~K (see, for example, 
\cite{shull1979,hollenbach1979,draine1993} for the structure of radiative shock).
Since at shock speeds $\simlt 200$~\kms, 
dust grains are mostly not destroyed although their size distribution may be  
modified by grain-grain shattering \citep{jones1996}, we expect 
dust grains in these hot and warm regions 
that will be mainly heated by attenuated X-ray and UV radiation 
and also by the recombination radiation from the ionized gas in those regions
\citep{hollenbach1979,andersen2011,lee2011,koo2014}. 
If there are enough gas and dust columns beyond these 
hot and warm regions, 
the dust IR emission could be further absorbed by cold dust 
and re-emitted at longer wavelengths. 
Therefore, we expect that, in these SNRs, a significant fraction of the shock energy 
might be converted into IR radiation of a broad spectral energy distribution  
at color temperature considerably lower than that of the 
collisionally heated dust. 
%Indeed, in Kes 17, for example, the  
%SED peaks at 100~$\mu$m and a two-temperature fit requires a dust component 
%at $\sim 27$ K \citep{lee2011}. 

%There could be some contribution 
%by collisional heating, but the temperature of dust grains will be 
%$\simlt 50$~K unless electron density is much higher than
%100 cm$^{-3}$ \citep{dwek2008}. 

%As an example, Figure 9 (left) shows 
%the dust (and gas) temperature profiles of radiative precursor 
%region for a 100 \kms\ shock propagating into a medium with 
%hydrogen density of 100 cm$^{-3}$. 
%The calculation was done by 
%using the plasma simulation code CLOUDY \citep[C13.03; ][]{ferland2013},
%and the incoming radiation field from the shock was obtained by using the 
%Raymond shock code \citep{raymond1979,cox1985}. 
%The temperature 
%of silicate (and graphite) grains of 0.01--0.1 $\mu$m radius ranges 
%20--40 K, and the resulting spectral energy distribution (SED) 
%of the FIR emission is broader than that of a single-temperature dust 
%component (Figure 9 right).
%The observed dust color temperatures (43--51 K) 
%are higher than this, and it could be due to 
%the emission from postshock dust 
%heated by both collisions and radiation \citep{andersen2011} 
%as well as the line radiation such as [O IV]+[Fe II] near 25.9 $\mu$m and 
%[O I] 63 $\mu$m.

%According to \citep{goncalves2011} 48-65 K,
%For comparison, Anderson, the temperatures are somewhat lower than this. but their 
%temperatures are from one to few slit observations so that could not 
%represent the entire SNR. We keep this in mind. 

It would be interesting to determine what fraction of shock energy is 
converted into the IR emission in these SNRs. We define 
the characteristic efficiency of converting shock energy into IR radiation 
($\effir$) for an SNR by the following equation
\begin{equation}  
\lir = {1\over 2}  \effir n_a \muH {v_s}^3 \times 4 \pi {R_s}^2 
=  3.7 \times 10^4 \effir n_{a,1} v_{s,2}^3 R_{s,1}^2~~~\lsun 
%=  3.66 \times 10^4 \effir n_{a,1} v_{s,2}^3 R_{s,1}^2~~~\lsun 
\label{eq-5}
\end{equation}
where $\lir$ is the IR luminosity, $n_a$ is the hydrogen nuclei number density 
of the ambient medium, $v_s$ and $R_s$ are the 
shock speed and radius of the SNR, and   
$\muH =2.34\times10^{-24}$ g
is the mean mass per hydrogen nucleus for a gas of cosmic abundance.
In the second equation, we normalized the parameters by their typical values, 
i.e., $n_{a,1}\equiv n_a/(10 {\rm \ cm}^{-3}$), $v_{s,2}\equiv v_s/(10^2 {\rm \ km~s}^{-1}$), 
and $R_{s,1}\equiv R_s/(10 {\rm \ pc}$). 
Note that ${1\over 2}  n_a \muH {v_s}^3$ is the incoming 
energy flux, so $\effir$ would be the ratio of IR luminosity to the 
energy incoming rate  
for a spherically symmetric SNR of radius $R_s$ in a uniform 
medium with a density of $n_a$. 
Of course real SNRs are not spherically symmetric, and   
$n_a$ and $v_s$ usually vary over the remnant. 
For radiative portion of the SNRs, 
we expect $n_a \simgt 1$~cm$^{-3}$ and $v_s\simlt 200$~\kms,
so the cooling time, 
$\tau_{\rm cool} \approx 3.8\times 10^3 v_{s,2}^{3.4}/n_a$~yr \citep{draine2011},
becomes shorter than the age of the SNR. 
We may obtain a constraint on $n_a$ and $v_s$ from X-ray observation 
by assuming that there is a pressure equilibrium between the shock 
pressure and the thermal pressure of the X-ray emitting gas, i.e., 
$n_a \mu_{\rm H} v_s^2 \approx 1.9n_e k_B T_e$ 
where $n_e$ and $T_e$ are the electron density 
and temperature of hot gas, respectively. 
(We assume fully ionized gas with $n(\text{He})/n(\text{H})=0.1$.)
The shock speed then is given by
\begin{equation}
v_{s,2}\approx 1.06 n_{a,1}^{-0.5}(\nefv T_7)^{0.5}f_V^{-0.25},
\label{eq-6}
\end{equation}
where $f_V$ is the volume filling factor of the X-ray emitting gas
and $T_7\equiv (T_e/10^7~{\rm K})$. 
Note that we used $\nefv$ (instead of $n_e$) because it is the parameter 
obtained from X-ray observation, e.g., X-ray flux is proportional to 
$\int n_e^2 dV=n_e^2 f_V (4\pi/3)R_s^3$. 
If we substitute this shock speed into Equation (5), the IR 
luminosity is given by
\begin{equation}  
\lir =  4.4 \times 10^4 \effir n_{a,1}^{-0.5} 
{\nefv T_7}^{1.5} f_V^{-0.75} R_{s,1}^2~~~\lsun 
%=  4.36 \times 10^4 
\label{eq-7}
\end{equation}
In the above equation, $\lir$ is from the IR observation (Table 1) and 
$\nefv$ and $T_7$ are from the X-ray observation (Table 2), so that if we know $f_V$ 
and $n_a$ we can determine $\effir$ (assuming that we know the distance 
to the SNR and therefore $R_s$). 

Table 4 summarizes the physical parameters of six SNRs with high IRX ratios 
and their $\effir$. We fixed $f_V=0.5$ because  
the volume filling factors are uncertain, although it has been estimated for 
some SNRs, e.g., 0.36 \citep[W49B;][]{kawasaki2005} and  
$\sim 0.4$ \citep[IC 443;][]{troja2006}. 
The ambient density is also uncertain.
\cite{chevalier1999} showed that some of these SNRs, i.e.,
W44, IC 443, and 3C391, are well described by an SNR model 
expanding within a molecular cloud with an 
interclump density of 5--25 cm$^{-3}$, while 
\cite{park2013} derived a somewhat lower density ($1.9$~cm$^{-3}$) for W44
from \schi\ 21~cm line observations \citep[see also ][]{koo1995}.
We adopted 3--15~cm$^{-3}$ for $n_a$ (see note in Table 4). 
It is not unusual to observe both atomic and molecular shocks 
in these SNRs, so the $n_a$ values in Table 4 should be considered 
as reference values. Note that we can scale the result 
using $\effir\propto n_a^{0.5}f_V^{0.75}$ (Equation 7).
Although there are considerable uncertainties 
in $\effir$ for a particular SNR, 
our result shows that $\effir\simgt 30$\% in general.
We can imagine that this `global' efficiency $\effir$ will be proportional to 
the fraction of surface area encountering dense ambient medium 
and thus the local efficiency of converting shock energy into IR radiation 
could be higher. 
For comparison, the fraction of shock energy that goes into 
the collisional-heating of dust is 1\%--10\% for shocks 
with speeds of 100--200~\kms \citep{draine1981}. 
Therefore, as we expect from their large IRX flux ratios, 
the IR luminosities of these SNRs are significantly greater than 
those from collisionally-heated dust. 

\subsection{LMC Supernova Remnants with High and Low IRX Flux Ratios}

Recently \cite{seok2015} conducted a systematic study of the LMC SNRs using 
\sst\ and {\em Chandra} data, and showed that the IRX flux ratios of the 
LMC SNRs are systematically lower than those of the Galactic SNRs, which 
is consistent with the low DGR and the low metallicity of the LMC, i.e.,  
the dust cooling rate is proportional to the DGR which is 
one-fourth of the Galactic value while
the gas cooling rate is proportional to metallicity which is 
one-half of the Galactic value.
They further classified the SNRs into three groups, 
i.e., SNRs with definite, partial, and 
lack of IRX correlation, and found 
that the IRX flux ratios of the SNRs with a partial
or lack of IRX morphological correlation  
are systematically higher than the ones with definite correlation.

In Figure 9, we plot the 
\irxobs\ values of the LMC SNRs against their dust temperatures. 
The IRX flux ratios are from \cite{seok2015} while 
dust temperatures are derived here from the 24 and 70 $\mu$m fluxes in \cite{seok2013}.
We can see that the SNRs with a 
lack of IRX morphological correlation 
not only have large \irxobs\ values but also have low temperatures as we 
have found for the Galactic SNRs. Their temperature ranges 
from 45 to 56~K, the maximum of 
which is a little higher than the 51 K of the Galactic SNRs. 
However,  
the dust temperature of the SNRs with definite IRX morphological correlation 
extends over a broad range from 48 to 80 K 
(excluding 1987A, for which $T_d$ is 166 K).
There is one SNR (0453$-$68.5) with a definite correlation 
that has an exceptionally low ($\sim 49$ K) dust temperature. 
\cite{williams2006} derived a comparable 
dust temperature (40--55 K) for the northern shell of this SNR.
Its low $T_d$ could be due to low electron density 
combined with low electron temperature, e.g.,   
$0.76$ cm$^{-3}$ and $0.29$ keV \citep{williams2006}. 
Figure 9 also shows that about two-thirds of the LMC SNRs 
are below the dashed line, which confirms the conclusion of \cite{seok2015} that 
the IRX flux ratio of the 
LMC SNRs are systematically lower than those of the Galactic SNRs.
(Note that \cite{seok2015} used the X-ray flux from the Chandra SNR catalog, so that 
their IRX flux ratios are different from ours.) 
In Figure 9, however, the SNRs at low and high dust temperatures are not necessarily 
below the dashed line. 
This may be because, as we have found in the Galactic SNRs, 
the IR emission of the SNRs at low temperatures is dominated by radiatively-heated 
dust grains while the X-ray emission of the SNRs at high temperatures 
is dominated by SN ejecta, so that the low DGR and the low metallicity 
do not explicitly appear in the IRX flux ratio. 
Indeed the SNRs with high dust temperatures, i.e., N132D, N103B, 0519$-$69.0, and 
0509$-$67.5,
are all the youngest ($\simlt 3,000$~yr) SNRs in the LMC.
In conclusion, 
the location of SNRs in the (\irxobs, $T_d$) diagram reflects the SNR environment 
as well as their natural properties. 

\section{Summary and Conclusion}

FIR dust emission is a dominant emission for SNRs, 
and it has been pointed out that 
FIR cooling by collisionally-heated dust 
could be more important than the X-ray cooling 
by hot gas in SNRs over most of their lifetimes.
With the development of IR astronomy, 
this was confirmed by comparing the 
IR to X-ray fluxes. 
The observed IRX flux ratios \irxobs\, however, 
scatter over three orders of magnitudes, and 
with the \iras\ data of several arcminute resolution,
it was difficult to explore the true nature of the IR emission for most SNRs.
In this paper, we conducted a systematic study of 
the IR and X-ray properties of 20 SNRs using the data from 
recent space missions to reveal details of morphology. 
We first showed that SNRs have 
diverse relative IR and X-ray morphologies, from very similar to very different, 
and that their resemblance may be represented by 
the correlation parameter \rfifty\ (\S~2.3).
We then have derived IRX flux ratios (\irxobs) of the SNRs 
and found that there is a systematic dependence on \rfifty.
By comparing \irxobs\ to the expected IRX flux ratios (\irxcoll) for hot plasma 
where dust grains are collisionally heated, we explored 
how the IRX flux ratios and IRX morphology are related to 
the SNR environment and their natural properties.
We summarize our main results as follows:

\noindent
1. The \irxobs\ values of SNRs range from 
0.3 to 200. The SNRs with high negative \rfifty\ 
have systematically higher IRX flux ratios. 

\noindent
2. The \irxobs\ values of all SNRs except a few are all below the expected ratio for hot, dusty plasma 
of constant DGR in collisional ionization equilibrium. 
This can be attributed to 
dust destruction and NEI gas cooling behind the shock front in some SNRs, 
but for most SNRs \irxobs\ values 
are either considerably larger or smaller than those of \irxcoll. 

\noindent
3. Among the SNRs with \irxobs\ values significantly less than 
those of \irxcoll, the most prominent
are young SNRs such as Cas A, Kepler, and Tycho. In these SNRs, the   
X-ray emission is dominated by metal-rich SN ejecta and 
our \irxcoll\ values are not 
applicable. We consider that this might be also the case for other young 
SNRs with anticorrelated IRX morphology in this category. 
There are, however, also SNRs without apparent X-ray emission from SN ejecta. 
They exhibit good IRX morphological correlation.  
For these SNRs, a low DGR of the ambient medium seems to be a plausible explanation. 

\noindent
4. All SNRs with \irxobs\ values significantly greater than \irxcoll\ ones 
have large negative \rfifty, i.e., are with anticorrelated IRX morphology. 
The dust temperature of these SNRs is systematically lower than those of the other SNRs. 
These SNRs are known to be interacting with a 
dense environment, and the IR emission in   
these SNRs is probably from dust heated by shock radiation rather than by collisions.
The global conversion efficiency of shock energy to IR emission 
is estimated to be more than a few tens of percent.

Our results show that the IRX flux ratios together with dust temperature are useful 
indicators for the SNR environment as well as their natural properties. 
We may infer that the SNRs with small \irxobs\ and high $T_d$ values are young 
SNRs with X-ray emission dominated by SN ejecta, whereas the SNRs with 
large \irxobs\ and low $T_d$ are evolved SNRs interacting with a dense environment. 
This was confirmed for the SNRs in the LMC. 
Such information might be particularly useful 
for the study of SNRs in galaxies where the SNRs are not resolved. 

\acknowledgements

We thank the anonymous referee and Jeonghee Rho for their comments 
which helped to improve the manuscript. This research was supported by the Basic Science Research Program through the National Research Foundation of Korea(NRF) funded by the Ministry of Science, 
ICT and Future Planning (2014R1A2A2A01002811). H.-J. K. was supported by NRF(National Research Foundation of Korea) Grant funded by the Korean Government (NRF-2012-Fostering Core Leaders of the Future Basic Science Program).
This work is based on observations made with the {\it Spitzer Space Telescope}, which is operated by the Jet Propulsion Laboratory, California Institute of Technology under a contract with NASA, data obtained from the {\it Chandra} Data Archive, and observations obtained with {\it XMM-Newton}, an ESA science mission with instruments and contributions directly funded by ESA Member States and NASA. This publication makes use of data products from the {\it Wide-field Infrared Survey Explorer}, which is a joint project of the University of California, Los Angeles, and the Jet Propulsion Laboratory/California Institute of Technology, funded by the National Aeronautics and Space Administration and the {\it ROSAT} Data Archive of the Max-Planck-Institut für extraterrestrische Physik (MPE) at Garching, Germany.

{}

\begin{figure}
\epsscale{0.7}
\plotone{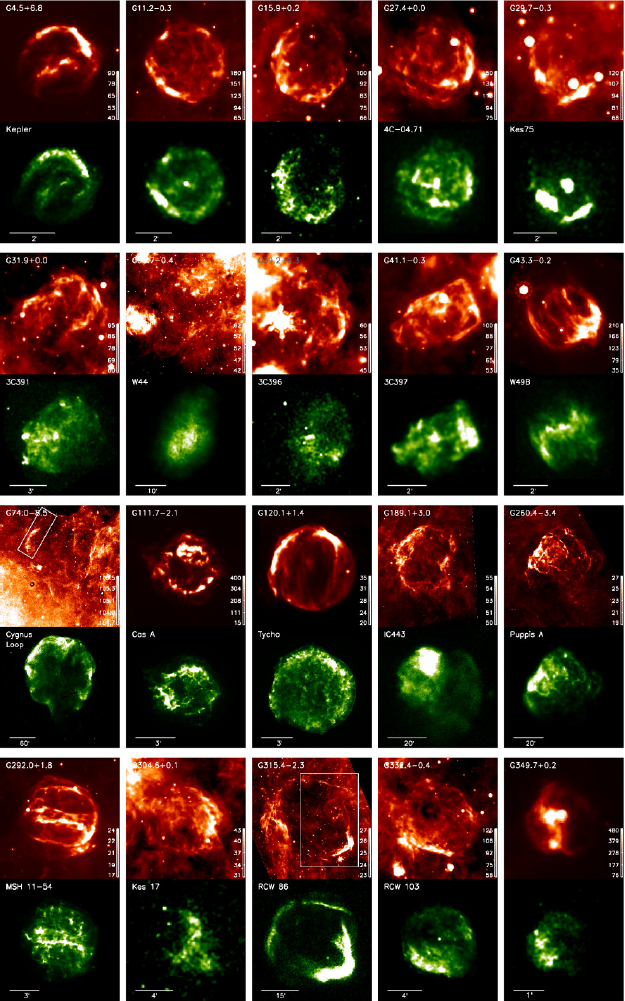}
 \caption{
IR and X-ray images of the 20 SNRs in Table 1. 
Red images in upper rows are \sst\ 24 $\mu$m images; 
green images in lower rows are \chandra\  
0.3--2.1 keV images from the \chandra\ SNR catalog
(http://hea-www.cfa.harvard.edu/ChandraSNR/)
except for the following:
IR image of Cygnus Loop is a \wise\ 22~$\mu$m image; 
X-ray images of Cygnus Loop, IC 443, and Puppis A 
are {\it ROSAT} PSPC 0.1--2.4 keV images, and 
X-ray images of Kes 17 and RCW 86 are 
\xmm-{\it Newton} images.
The color intensity scale is linear in both IR and X-ray images.
The white boxes in some SNR images represent areas where the correlation 
coefficient between IR and X-ray intensities 
in \S~2.3 are derived.
%The SNR names are labeled in individual frames. 
%The angular scale is given in each frame by a scale bar. 
}
\label{fig1}
\end{figure}

\begin{figure}
\epsscale{1.5}
\plotone{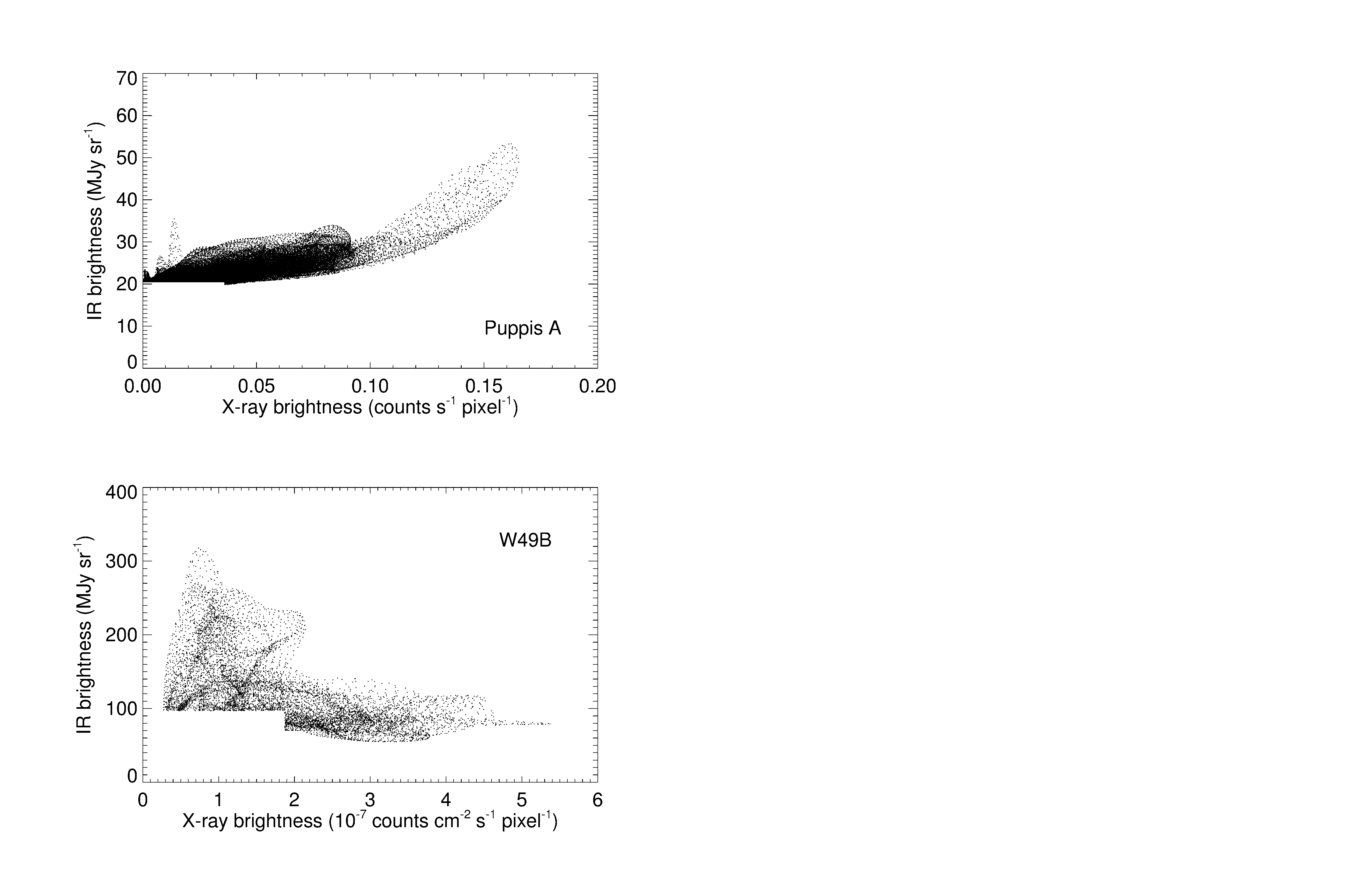}
 \caption{
Scattered diagrams of 24 $\mu$m versus X-ray brightness   
for Puppis A (upper frame) and W49B (lower frame) 
that have the largest positive (0.76) and the largest negative 
($-0.49$) correlation coefficient \rfifty, respectively.
}
\label{fig2}
\end{figure}

\begin{figure}
\epsscale{0.8}
\plotone{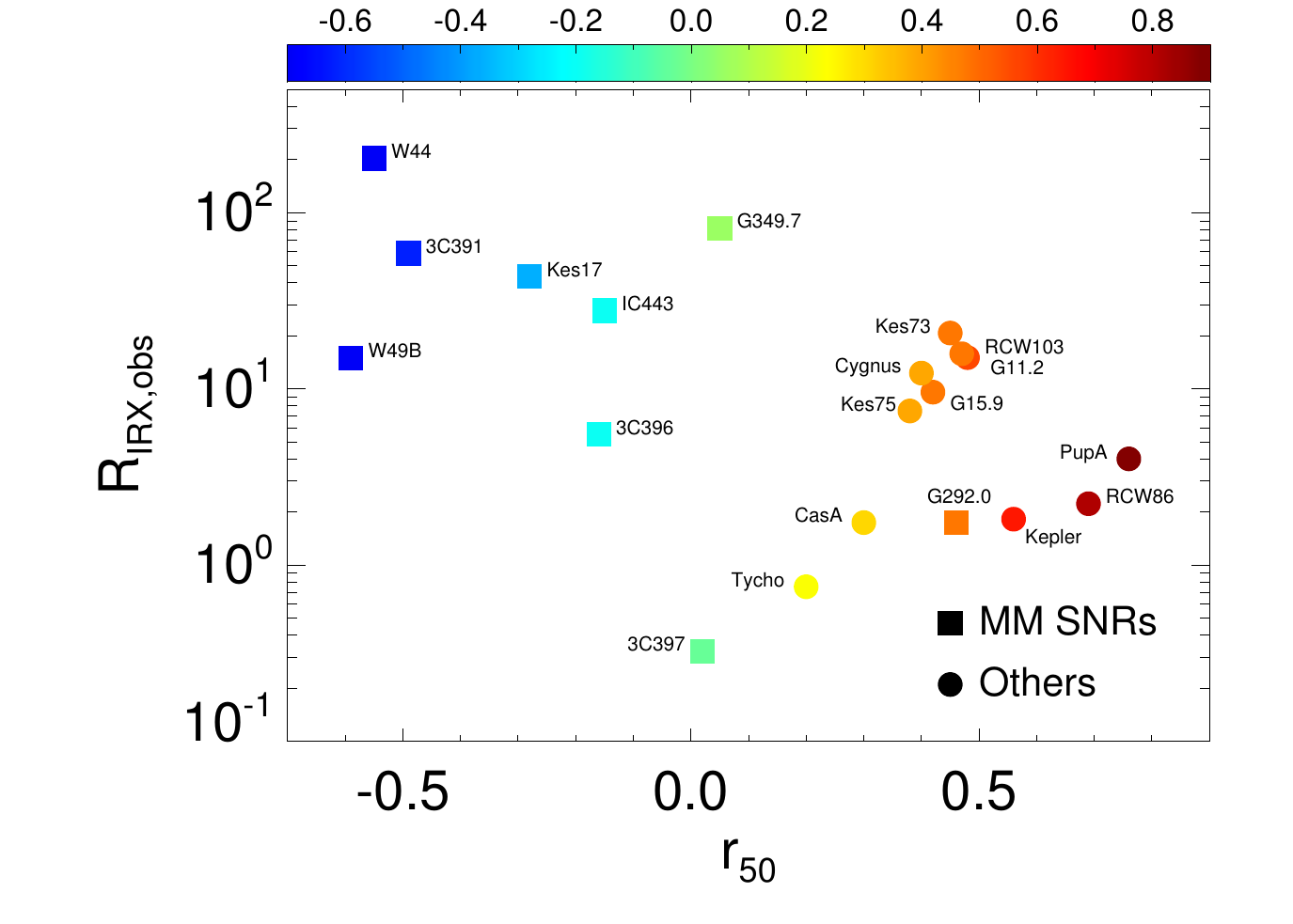}
 \caption{
Observed IRX flux ratio versus \rfifty.
The color of SNR symbols varies from blue to red with $r_{50}$ increasing 
from $-0.7$ to 0.9 as in the color bar. 
The square symbol indicates that the SNR is an MM-type SNR.
}
\label{fig3}
\end{figure}

\newpage

\begin{figure}
\epsscale{1.0}
\plotone{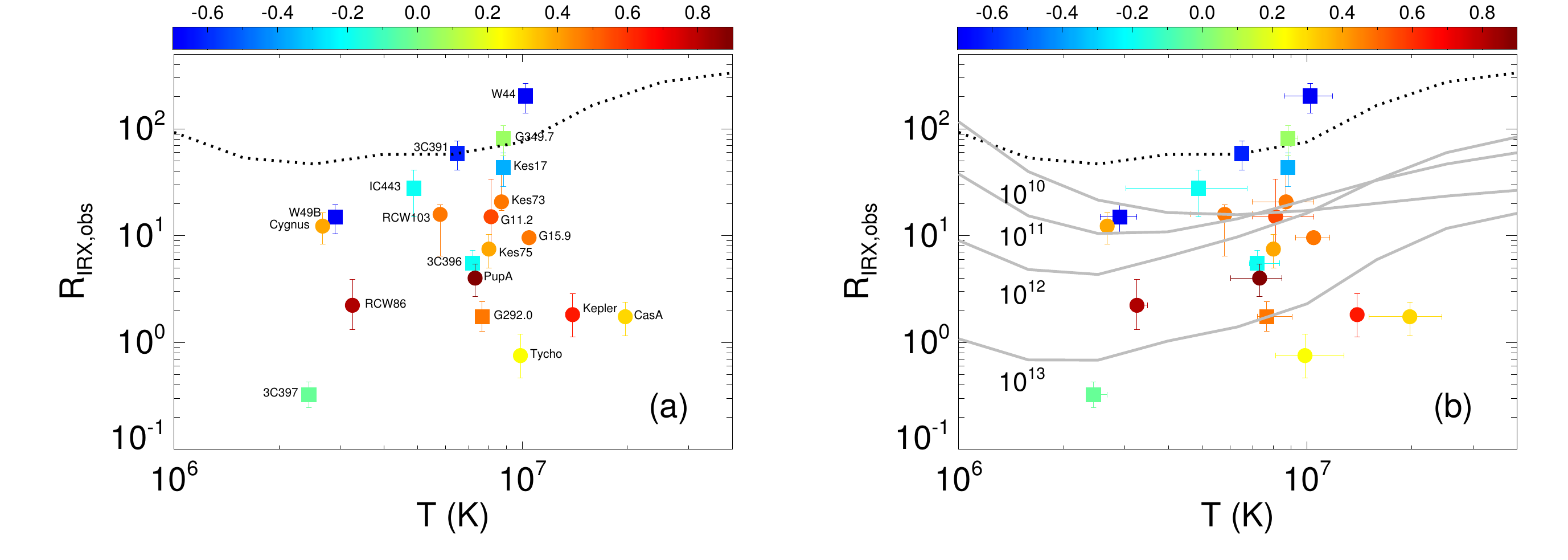}
 \caption{
(a) Observed IRX flux ratio versus temperature of X-ray emitting plasma. 
The dotted line is a theoretical curve representing the IRX flux ratio of 
hot plasma in thermal and collisional ionization equilibrium. 
The colors and symbols are the same as in Figure~\ref{fig3}.
(b)  Same as (a) but with 
time-dependent theoretical curves overlaid. 
The solid curves represent the IRX flux ratios for hot plasma 
at $\tau_{\rm ion}=10^{10}, 10^{11}, 10^{12}$, and $10^{13}$ 
cm$^{-3}$ s after being shocked (see text for details).
}
\label{fig4}
\end{figure}

\begin{figure}
\epsscale{1.0}
\plotone{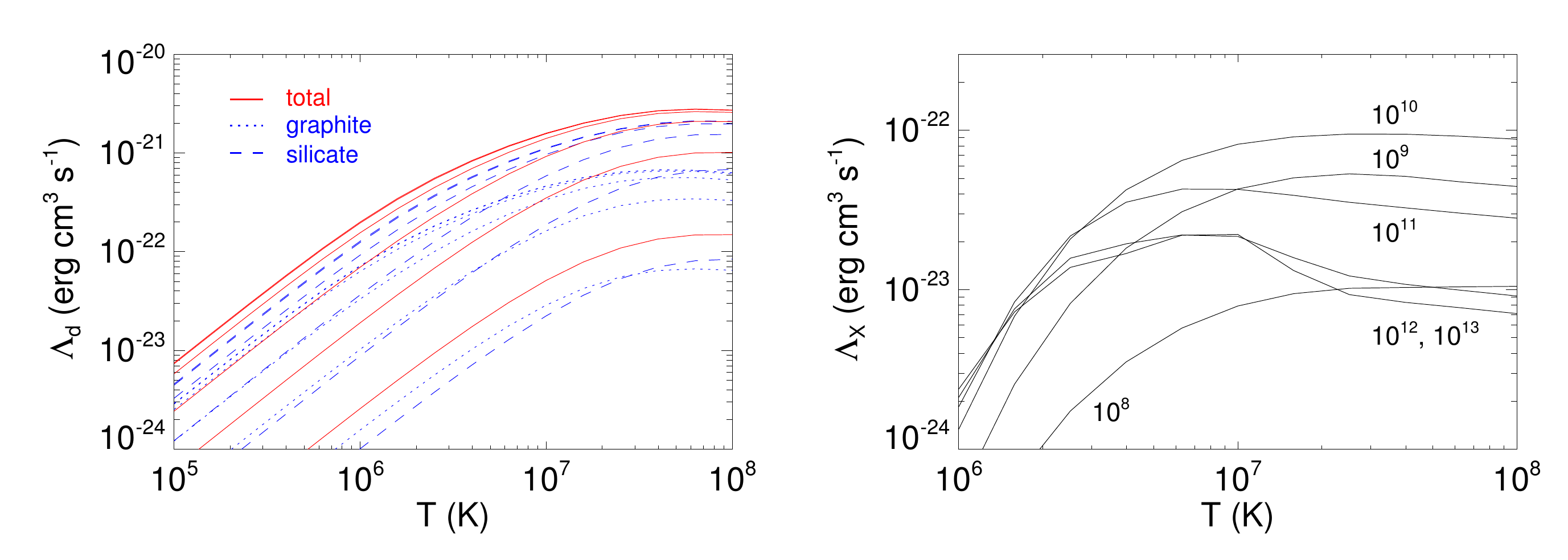}
 \caption{(Left) Time-dependent dust cooling function of shocked gas at  
$\tau_{\rm ion}=10^8, 10^9, 10^{10}, 10^{11}$, and $10^{12}$ 
cm$^{-3}$ s  from top to bottom.
(Right) Time-dependent NEI X-ray (0.3--2.1 keV) cooling function.
The curves are labeled by \tauion\ (cm$^{-3}$ s). 
}
\label{fig5}
\end{figure}

\begin{figure}
\epsscale{0.8}
\plotone{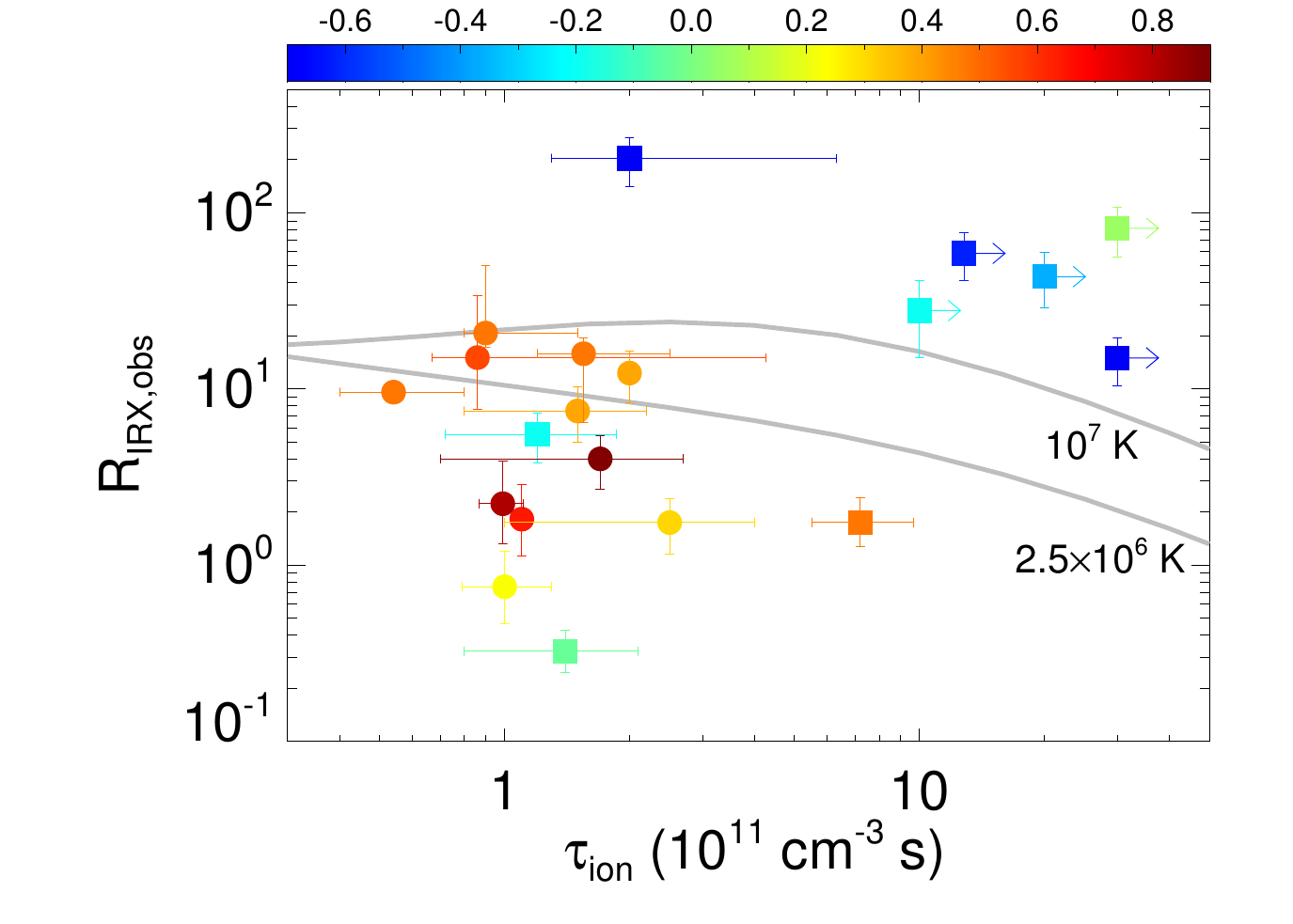}
 \caption{
Observed IRX flux ratio versus ionization timescale.
Solid curves are theoretical curves for hot plasma 
at $T_e=10^7$ K and $2.5\times 10^6$~K.
The colors and symbols are the same as in Figure~\ref{fig3}.
}
\label{fig6}
\end{figure}

\begin{figure}
\epsscale{0.8}
\plotone{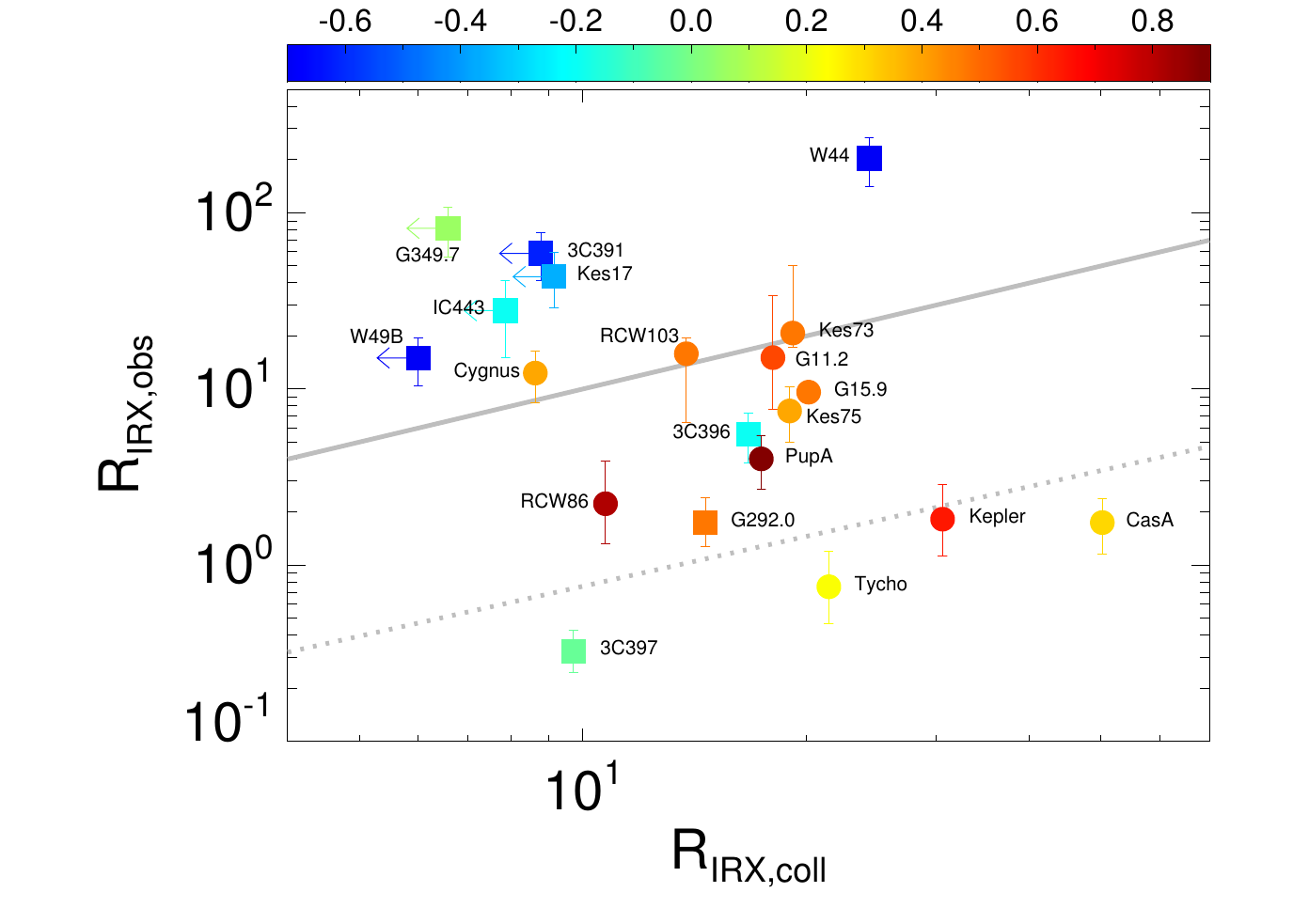}
 \caption{
Observed IRX flux ratio versus theoretical IRX flux ratio for 
hot plasma where dust grains are heated by collisions. 
The solid and dotted curves represent when 
$R_{\rm IRX,obs}=R_{\rm IRX,coll}$ and $0.1R_{\rm IRX,coll}$, respectively. 
The errors in \irxcoll\ are not shown but may be found in Table 3.
The colors and symbols are the same as in Figure~\ref{fig3}.
}
\label{fig7}
\end{figure}

\begin{figure}
\epsscale{0.8}
\plotone{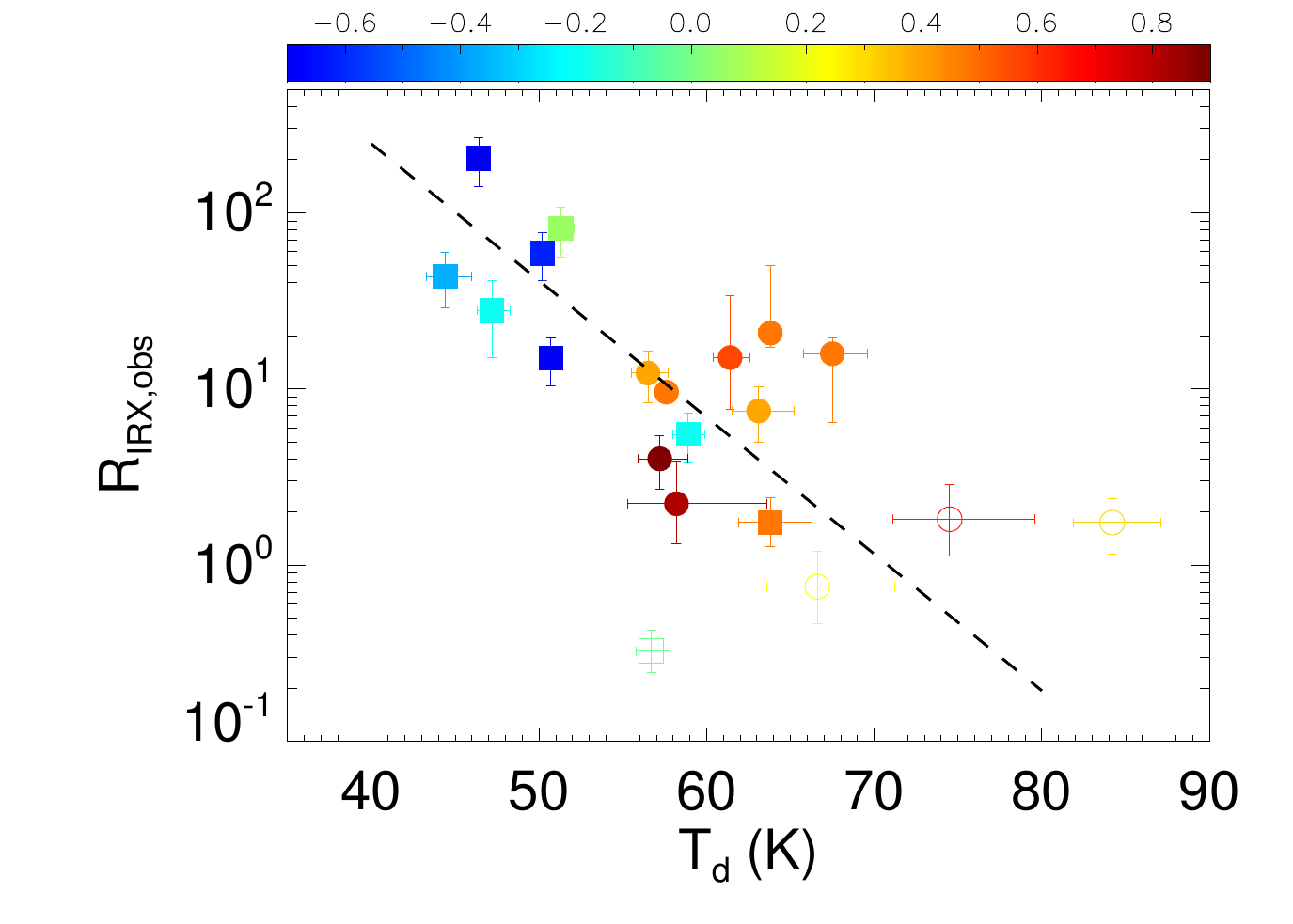}
 \caption{
Observed IRX flux ratio versus dust temperature $T_d$. 
The dashed line is the best-fit curve obtained by least-squares fitting 
(Equation 4). The empty circles represent the SNRs with low 
\irxobs\ values in Figure 7, which are 
excluded in the fit. 
The colors and the other symbols are the same as in Figure~\ref{fig3}.
}
\label{fig8}
\end{figure}

\begin{figure}
\epsscale{0.8}
\plotone{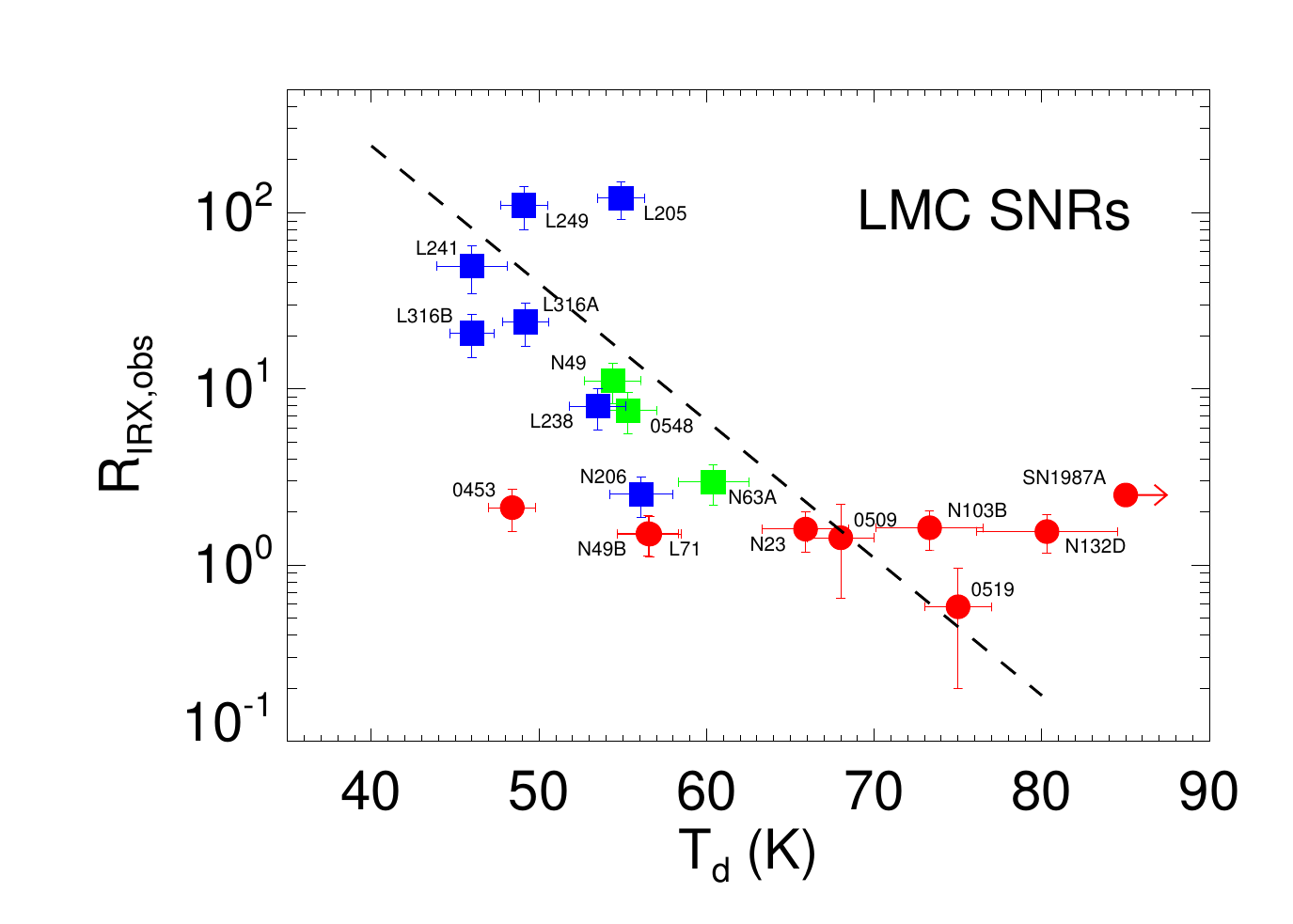}
 \caption{
Observed IRX flux ratio versus dust temperature $T_d$ for the SNRs in the LMC.
The red circle, green box, and blue box 
symbols represent the SNRs with definite (red circle), partial (green box), and 
lack of (blue box) IRX morphological correlation, respectively.
The dashed line is the best-fit curve for the Galactic SNRs in 
Figure 8.
}
\label{fig9}
\end{figure}

\clearpage

\begin{deluxetable}{llccccc}
\tabletypesize{\scriptsize}
%\rotate
\tablecaption{Infrared Fluxes and Dust Temperatures of 20 SNRs.\label{tbl-1}}
\tablewidth{0pt}
\tablecolumns{7} \tablehead{
\colhead{SNR} &\colhead{Other Name} & \colhead{$F_{\rm 24}$} & \colhead{$F_{\rm 70}$} &
\colhead{$T_{\rm d}$} & \colhead{$F_{\rm IR}$} & \colhead{Reference} \\
&& \colhead{(Jy)} & \colhead{(Jy)} & \colhead{(K)} & \colhead{($10^{-9}$ ergs cm$^{-2}$ s$^{-1}$)} &
}
\startdata
   G4.5$+$6.8 &       Kepler &   9.50 ( 1.0) &   10.2 ( 2.7) &   75.0 (-3.5,+5.2) &   1.83 (-0.45,+0.91) & 1 \\
  G11.2$-$0.3 &      \nodata &   38.6 ( 1.5) &  124.7 (14.0) &   61.6 (-1.1,+1.2) &   12.8 ( -1.3,+ 1.6) & This work \\
  G15.9$+$0.2 &      \nodata &   10.1 ( 0.4) &   49.1 ( 2.7) &   57.8 (-0.5,+0.6) &   4.32 (-0.24,+0.27) & This work \\
  G27.4$+$0.0 &       Kes 73 &   30.1 ( 1.2) &   76.6 ( 4.1) &   64.1 (-0.7,+0.7) &   8.73 (-0.53,+0.61) & This work \\
  G29.7$-$0.3 &       Kes 75 &   10.2 ( 0.4) &   27.9 ( 4.9) &   63.3 (-1.6,+2.1) &   3.08 (-0.45,+0.66) & This work \\
  G31.9$+$0.0 &        3C391 &   33.9 ( 1.4) &  434.9 (21.9) &   50.3 (-0.4,+0.4) &   28.6 ( -1.3,+ 1.5) & This work \\
  G34.7$-$0.4 &          W44 &   94.4 ( 9.5) &  2243. (114.) &   46.5 (-0.6,+0.7) &  129.2 ( -9.2,+11.3) & This work \\
  G39.2$-$0.3 &        3C396 &    6.4 ( 0.6) &   26.8 ( 1.6) &   59.1 (-0.9,+1.1) &   2.49 (-0.22,+0.28) & This work \\
  G41.1$-$0.3 &        3C397 &   18.8 ( 0.8) &  101.4 (11.9) &   56.9 (-0.9,+1.1) &   8.60 (-0.82,+1.06) & This work \\
  G43.3$-$0.2 &         W49B &   62.8 ( 2.5) &   750. ( 38.) &   50.8 (-0.4,+0.4) &   50.2 ( -2.3,+ 2.6) & This work \\
  G74.0$-$8.5 &  Cygnus Loop &   172. ( 17.) &  1430. (140.) &   47.5 (-1.3,+1.7) &   85.3 (-12.6,+20.4) & 2 \\
 G111.7$-$2.1 &        Cas A &   202. ( 20.) &   120. ( 12.) &   85.0 (-2.4,+3.0) &   32.1 ( -5.1,+ 7.3) & 3, 4\\
 G120.1$+$1.4 &        Tycho &   19.5 ( 3.0) &   38.5 (10.8) &   66.9 (-3.0,+4.7) &   4.94 (-1.20,+2.50) & 1 \\
 G189.1$+$3.0 &       IC 443 &   68.8 ( 6.9) &  1349. (202.) &   47.6 (-0.9,+1.2) &   80.7 ( -8.9,+12.5) & This work \\
 G260.4$-$3.4 &     Puppis A &   187. ( 19.) &   955. (143.) &   56.8 (-1.3,+1.7) &   82.9 (-10.8,+15.6) & This work \\
 G292.0$+$1.8 &  MSH 11$-$54 &   12.0 ( 0.7) &   30.6 ( 6.1) &   64.0 (-1.9,+2.5) &   3.48 (-0.58,+0.91) & 5 \\
 G304.6$+$0.1 &       Kes 17 &    2.2 ( 0.4) &    75. ( 15.) &   44.4 (-1.1,+1.6) &   4.08 (-0.59,+0.96) & 6\\
 G315.4$-$2.3 &       RCW 86 &   0.69 (0.19) &   3.10 (1.06) &   58.5 (-3.0,+5.5) &   0.28 (-0.08,+0.20) & 7\\
 G332.4$-$0.4 &      RCW 103 &   75.0 ( 7.5) &  137.0 (17.2) &   67.9 (-1.7,+2.1) &   18.3 ( -2.6,+ 3.7) & This work \\
 G349.7$+0.2$ &      \nodata &   39.6 ( 3.9) &  435.0 (21.8) &   51.4 (-0.7,+0.8) &   29.8 ( -2.3,+ 2.8) & This work \\

\enddata
\tablecomments{$F_{\rm 24}$ and $F_{\rm 70}$ are \spitzer\ MIPS 24 and \herschel\ PACS 70 $\mu$m fluxes except for the following sources: 
G74.0$-$8.5 ({\em IRAS} 25 and 60 $\mu$m fluxes); 
G189.1$+$3.0, G260.4$-$3.4, G315.4$-$2.3 (\spitzer\ MIPS 70 $\mu$m flux); 
G292.0$+$1.8, G304.6$+$0.1 (AKARI 65 $\mu$m flux). $T_{\rm d}$ is 
the color temperature for 
thermal dust emission, and $F_{\rm IR}$ is the integrated infrared flux.
The numbers in parenthesis are $1\sigma$ errors.
}

\tablecomments{$F_{\rm 24}$ and $F_{\rm 70}$ are {\em measured} (i.e., not color-corrected), background-subtracted
fluxes of the {\em entire} SNR except the following sources:\\
G4.5$+$6.8: Flux of warm dust component in Tables 2 and 4 of \cite{gomez2012}. \\
G11.2$-$0.3: $F_{\rm 70}$ is a bootstrapped flux from the 70 $\mu$m flux of the SNR shell
             (excluding the central area) using its $F_{\rm 70}/F_{\rm 24}$ ratio. \\
G29.7$-$0.3: $F_{\rm 70}$ is a bootstrapped flux from the 70 $\mu$m flux of the bright shell
             using its $F_{\rm 70}/F_{\rm 24}$ ratio. \\
G41.1$-$0.3: $F_{\rm 70}$ is a bootstrapped flux from the 70 $\mu$m flux of the bright eastern shell
             using its $F_{\rm 70}/F_{\rm 24}$ ratio. \\
G111.7$-$2.1: $F_{\rm 24}$ is from \cite{hines2004} and $F_{\rm 70}$ is the
             flux of the warm dust component in Table 1 of \cite{barlow2010}. \\
G120.1$+$1.4: Flux of warm dust component in Tables 3 and 4 of \cite{gomez2012}. \\
G304.6$+$0.1: Flux of the western shell in Table 4 of \cite{lee2011}.
This is essentially equal to the total flux at wavelengths $\ge 60$ $\mu$m.\\
G315.4$-$2.3: Flux of the NW region in Table 2 of \cite{williams2011}. This is a small fraction of the total flux.\\
G332.4$-$0.4: $F_{\rm 70}$ is a bootstrapped flux from the 70 $\mu$m flux of the bright southern shell
             using its $F_{\rm 70}/F_{\rm 24}$ ratio. \\
G349.7$+$0.2: Flux excluding the central bright source.
}

\tablerefs{References for 24 and 70 $\mu$m fluxes: 
(1) \cite{gomez2012}; (2) \cite{braun1986}; (3) \cite{hines2004}; (4) \cite{barlow2010}; 
(5) \cite{lee2009}; (6) \cite{lee2011}; (7) \cite{williams2011}
}
%\tablenotetext{d}{This is a complicated area, but the shell is clearly visible, particularly at 70 um. We just derive the total flux using a thin shell surrounding the SNR as the background area. About half of the flux 
%      is contributed by the northern bright ring-like structure, which might be associated with the remnant.
%      southern shell: 1.2 (0.1) 39.8 (2.1)
%      northern bright shell: 51.7 (5.2)  1298.2 (65.0)}

\end{deluxetable}

\begin{deluxetable}{llllllc}
\tabletypesize{\scriptsize}
%\rotate
\tablecaption{X-ray Parameters of 20 SNRs.\label{tbl-2}}
\tablewidth{0pt}
\tablecolumns{7} \tablehead{
\colhead{SNR} &\colhead{Other Name} & \colhead{$N({\rm H})$} & \colhead{$T_e$} &
\colhead{$\tau_{\rm ion}$} & \colhead{$F_X$} & \colhead{Reference} \\
&& \colhead{($10^{22}$~cm$^{-2}$)} & \colhead{(keV)} & \colhead ($10^{11}$ cm$^{-3}$ s) &
 \colhead{($10^{-9}$ ergs cm$^{-2}$ s$^{-1}$)} &
}
\startdata
   G4.5$+$6.8 &       Kepler &   0.52 (  ...,  ...) &  1.2  (  ...,  ...) &  1.1  (  ...,  ...) &   1.02 (  ...,  ...) &  2, 2, 2, 1 \\
  G11.2$-$0.3 &      \nodata &   2.2  ( -0.4,  0.6) &  0.7  ( -0.1,  0.2) &  0.86 (-0.19,  3.4) &   0.87 (-0.42, 1.08) &  1, 1, 1, 1 \\
  G15.9$+$0.2 &      \nodata &   3.9  ( -0.2,  0.2) &  0.9  ( -0.1,  0.1) &  0.54 (-0.14, 0.26) &   0.46 (-0.01, 0.06) &  3, 3, 3, 1 \\
  G27.4$+$0.0 &       Kes 73 &   2.35 (-0.15, 0.65) &  0.75 (-0.15, 0.15) &  0.90 ( -0.1,  0.6) &   0.43 (-0.07, 0.61) &  1, 1, 1, 1 \\
  G29.7$-$0.3 &       Kes 75 &   3.96 (  ...,  ...) &  0.69 (-0.02, 0.01) &  1.5  ( -0.7,  0.7) &   0.42 (  ...,  ...) &  4, 4, 4, 4 \\
  G31.9$+$0.0 &        3C391 &   3.1  ( -0.1,  0.1) &  0.56 (-0.01, 0.01) & $\ge 12.8$          &   0.49 (  ...,  ...) &  5, 5, 5, 5  \\
  G34.7$-$0.4 &          W44 &   1.0  ( -0.2,  0.6) &  0.88 (-0.14, 0.14) &  2.0  (-0.7,   4.3) &   0.64 (  ...,  ...) &  6, 6, 6, 6 \\
  G39.2$-$0.3 &        3C396 &   4.65 (-0.26, 0.11) &  0.62 (-0.03,  0.1) &  1.20 (-0.48, 0.66) &   0.46 (  ...,  ...) &  7, 7, 7, 7  \\
  G41.1$-$0.3 &        3C397 &   3.05 ( -0.2,  0.2) &  0.21 (-0.01, 0.02) &  1.4  ( -0.6,  0.7) &  27.   (   -6,    8) &  8, 8, 8, 1 \\
  G43.3$-$0.2 &         W49B &   5.2  (  ...,  ...) &  0.25 (-0.03, 0.03) & $\ge 30$            &   3.40 (  ...,  ...) &  9, 9, 9, 9 \\
  G74.0$-$8.5 &  Cygnus Loop &   0.04 (  ...,  ...) &  0.23 (-0.01, 0.01) &  2.0  ( -0.1,  0.1) &   9.98 (  ...,  ...) & 10,10,10,11 \\
 G111.7$-$2.1 &        Cas A &   1.4  ( -0.5,  0.5) &  1.7  ( -0.4,  0.4) &  2.5  ( -1.5,  1.5) &  18.7  (  ...,  ...) & 12,13,13, 1 \\
 G120.1$+$1.4 &        Tycho &   0.70 (  ...,  ...) &  0.85 (-0.15, 0.25) &  1.00 (-0.21,  0.3) &   6.7  (  ...,  ...) & 14,15,15, 1 \\
 G189.1$+$3.0 &       IC 443 &   0.69 (-0.15, 0.15) &  0.42 (-0.16, 0.16) & $\ge 10$            &   2.9  ( -1.3,  1.3) & 16,16,16,16 \\
 G260.4$-$3.4 &     Puppis A &   0.24 ( -0.1, 0.11) &  0.63 (-0.11,  0.1) &  1.7  (   -1,    1) &  21.   (  ...,  ...) & 17,17,17,18 \\
 G292.0$+$1.8 &  MSH 11$-$54 &   0.54 (-0.04, 0.04) &  0.66 (-0.04, 0.12) &  7.2  ( -1.7,  2.5) &   2.04 (-0.44, 0.58) & 19,20,20, 1 \\
 G304.6$+$0.1 &       Kes 17 &   3.79 (-0.14, 0.03) &  0.76 (-0.01, 0.03) & $\ge 20$            &   0.09 (  ...,  ...) & 21,21,21,21 \\
 G315.4$-$2.3 &       RCW 86 &   0.63 (-0.05, 0.03) &  0.28 (-0.01, 0.02) &  0.99 (-0.12, 0.12) &   0.13 (  ...,  ...) & 22,22,22,22 \\
 G332.4$-$0.4 &      RCW 103 &   0.55 (-0.25, 0.15) &  0.50 ( -0.1,  0.1) &  1.55 (-0.35, 0.95) &   1.18 (-0.68, 0.13) &  1, 1, 1, 1 \\
 G349.7$+0.2$ &      \nodata &   7.10 ( -0.1,  0.1) &  0.76 (-0.03, 0.05) & $\ge 30$            &   0.37 (  ...,  ...) & 23,23,23,23 \\

\enddata
\tablecomments{$N({\rm H})$, $T_e$, $\tau_{\rm ion}$, and $F_X$ are hydrogen absorbing columns, electron temperature,
ionization timescale, and integrated (0.3--2.1 keV) X-ray flux, respectively.
The numbers in parenthesis are $1\sigma$ errors. $N({\rm H})$ without errors indicates that
it is fixed in the X-ray spectral analysis. For the other parameters, the numbers without errors indicate that
the errors are not given in the references.
When the CIE model is used for spectral analysis, we arbitrarily 
adopted $\tau_{\rm ion}\ge 3\times 10^{12}$~cm$^{-3}$ s.
}
\tablecomments{All the parameters including $F_X$ are for the {\em entire} SNRs except the following sources:\\
G29.7$-$0.3: $T_e$ and \tauion\ are average parameters of the diffuse and clump components in Table 4 of \cite{helfand2003}. \\
G41.1$-$0.3: $T_e$ and \tauion\ are average parameters of eastern and western lobes in Table 3 of \cite{safiharb2005}. \\
G43.3$-$0.2: $T_e$ and \tauion\ are those of the soft component in Table 2 of \cite{kawasaki2005}. \\
G189.1$+$3.0: Parameters are mean values of Shell A regions in Table 2 (see also Figure 7) of \cite{troja2006}.
              The flux is obtained by assuming a gas sphere of radius 7 pc with electron density 
              $1.6\pm 0.7$ cm$^{-3}$ and a volume filling factor of 0.5 \cite[cf.][]{troja2006}. \\
G260.4$-$3.4: Flux from \cite{dubner2013}; other parameters are mean values of the regions 
(excluding ejecta knots) in Table 3 of \cite{hwang2008}. \\
G292.0$+$1.8: $T_e$ and \tauion\ are those of the circumstellar medium (region 1) in Table 1 of \cite{park2004}. \\
G315.4$-$2.3: Parameters of the NW shell in Table 1 of \cite{williams2011}. \\
G349.7+0.2: Parameters of the soft component in Table 2 of \cite{lazendic2005}.
}
\tablerefs{References for $N({\rm H})$, $T_e$, \tauion, and $F_X$:
(1) this work; (2) \cite{burkey2013}; (3) \cite{reynolds2006}; (4) \cite{helfand2003};
(5) \cite{chen2004}; (6) \cite{harrus1997}; (7) \cite{harrus1999};
(8) \cite{safiharb2005}; (9) \cite{kawasaki2005}; (10) \cite{tsunemi2007}; (11) \cite{ku1984};
(12) \cite{lee2014}; (13) \cite{hwang2012}; (14) \cite{cassamchena2007};
(15) \cite{hwang1997};
(16) \cite{troja2006}; (17) \cite{hwang2008}; (18) \cite{dubner2013}; (19) \cite{lee2010};
(20) \cite{park2004}; (21) \cite{gelfand2013}; (22) \cite{williams2011};
(23) \cite{lazendic2005}
}
\end{deluxetable}

\begin{deluxetable}{llcrll}
\tabletypesize{\scriptsize}
%\rotate
\tablecaption{Infrared-X-ray Correlation Parameters of 20 SNRs.\label{tbl-3}}
\tablewidth{0pt}
\tablecolumns{6} \tablehead{
\colhead{SNR} &\colhead{Other Name} & \colhead{SN type (MM?)} & \colhead{$r_{50}$} &
\colhead{\irxobs} & \colhead{\irxcoll} 
}
\startdata

   G4.5$+$6.8 &       Kepler &     S &   0.56 &   1.82 (-0.70, 1.05) &   30.6 (  ...,  ...) \\
  G11.2$-$0.3 &      \nodata &     C &   0.48 &   15.1 ( -7.4, 18.8) &   18.0 ( -2.6,  6.2) \\
  G15.9$+$0.2 &      \nodata &     S &   0.42 &   9.59 (-0.56, 1.39) &   20.2 ( -2.4,  3.9) \\
  G27.4$+$0.0 &       Kes 73 &    S? &   0.45 &   20.8 ( -3.6, 29.5) &   19.2 ( -3.5,  5.2) \\
  G29.7$-$0.3 &       Kes 75 &     C &   0.38 &   7.50 (-2.49, 2.76) &   19.0 ( -1.7,  0.6) \\
  G31.9$+$0.0 &        3C391 &  S(y) &  -0.49 &  58.8 (-17.8, 17.9) &   $\le 8.78$  \\
  G34.7$-$0.4 &          W44 &  C(y) &  -0.55 &  203.8 (-62.8, 63.6) &   24.3 ( -6.8,  3.0) \\
  G39.2$-$0.3 &        3C396 &  C(y) &  -0.16 &   5.52 (-1.72, 1.76) &   16.7 ( -1.4,  3.3) \\
  G41.1$-$0.3 &        3C397 &  S(y) &   0.02 &   0.32 (-0.08, 0.10) &   9.72 (-1.21, 2.55) \\
  G43.3$-$0.2 &         W49B &  S(y) &  -0.59 &  15.0 ( -4.6,  4.6) &   $\le 2.41$  \\
  G74.0$-$8.5 &  Cygnus Loop &     S &   0.40 &   12.3 ( -3.9,  4.1) &   8.63 (-0.27, 0.26) \\
 G111.7$-$2.1 &        Cas A &     S &   0.30 &   1.74 (-0.59, 0.65) &   50.2 (-18.4, 13.7) \\
 G120.1$+$1.4 &        Tycho &     S &   0.20 &   0.75 (-0.29, 0.44) &   21.5 ( -3.6,  7.9) \\
 G189.1$+$3.0 &       IC 443 &  C(y) &  -0.15 &   27.8 (-12.8, 13.2) &   $\le 7.87$   \\
 G260.4$-$3.4 &     Puppis A &     S &   0.76 &   4.01 (-1.31, 1.42) &   17.4 ( -3.5,  3.0) \\
 G292.0$+$1.8 &    MSH 11-54 &  S(y) &   0.46 &   1.75 (-0.47, 0.67) &   14.6 ( -2.8,  4.6) \\
 G304.6$+$0.1 &       Kes 17 &  S(y) &  -0.28 &   43.4 (-14.5, 16.5) &   $\le 9.15$  \\
 G315.4$-$2.3 &       RCW 86 &     S &   0.69 &   2.23 (-0.91, 1.68) &   10.7 ( -0.3,  0.3) \\
 G332.4$-$0.4 &      RCW 103 &     S &   0.47 &   15.8 ( -9.4,  3.6) &   13.8 ( -2.6,  2.7) \\
 G349.7$+0.2$ &      \nodata &  S(y) &   0.05 &   81.6 (-25.3, 25.7) &   $\le 6.58$ \\
\enddata
\tablecomments{The `(y)' in SN type indicates that the SNR
is of mixed-morphology, i.e., shell/composite-type in radio with
center-filled thermal X-rays. See text for explanations of the other parameters.}
\end{deluxetable}

\begin{deluxetable}{lrrllrllll}
\tabletypesize{\scriptsize}
%\rotate
\tablecaption{Shock Energy Conversion Efficiency of SNRs with Large IRX Flux Ratios\label{tbl-4}}
\tablewidth{0pt}
\tablecolumns{10} \tablehead{
\colhead{SNR} & \colhead{$d$} & \colhead{$R_s$} &
\colhead{$L_X$} & \colhead{$T_e$} & \colhead{$n_e\sqrt{f_V}$} & \colhead{$L_{\rm IR}$} &
\colhead{$n_a$} & \colhead{$v_s$} & \colhead{$\epsilon_{\rm IR}$} \\
& \colhead{(kpc)} & \colhead{(pc)} & \colhead {($L_{\odot}$)} &
 \colhead{($10^{7}$ K)} & \colhead{(cm$^{-3}$)} & \colhead {($L_{\odot}$)}  &
 \colhead{(cm$^{-3}$)} & \colhead{(km s${^-1}$)} &  
}
\startdata
     3C391 &    8.0 &    7.5 &    9.8e+02 &   0.65 &   2.04 &    7.5e+04 &     10 &    144 &   0.76 \\
       W44 &    3.0 &   11.0 &    1.8e+02 &   1.02 &   0.43 &    5.7e+04 &      3 &    152 &   0.64 \\
      W49B &   11.4 &    6.6 &    1.4e+04 &   0.29 &  12.23 &    2.6e+05 &     10 &    237 &   0.80 \\
    IC 443 &    1.5 &    7.0 &    2.0e+02 &   0.49 &   1.14 &    1.5e+04 &     15 &     76 &   0.39 \\
     Kes17 &   10.0 &   10.0 &    2.8e+02 &   0.88 &   0.69 &    4.2e+04 &     10 &     98 &   0.24 \\
G349.7+0.2 &   22.0 &    6.4 &    5.6e+03 &   0.88 &   6.04 &    4.4e+05 &     10 &    290 &   1.0 \\
\enddata
\tablecomments{The conversion efficiency of shock energy to dust IR radiation $\epsilon_{\rm IR}$ 
can be derived from the observed 0.3--2.1 keV X-ray ($L_X$) and IR ($L_{\rm IR}$) luminosities, but 
depends on two uncertain parameters; the ambient density of H nuclei $n_a$ and the volume filling factor of X-ray emitting gas 
$f_V$ (see Equation 7). We fixed $f_V=0.5$ and assumed $n_a=10$ cm$^{-3}$ 
except for the following sources:\\
W44: \cite{park2013} obtained a mean density of 1.9 cm$^{-3}$ and shocked speed 
     of 135~\kms\ from \schi\ observations. \cite{chevalier1999} reviewed available
     observations and concluded that $n_a=4$--5 cm$^{-3}$ and $v_s\sim 150$~\kms. 
     We adopt $n_a=3$ cm$^{-3}$. \\
IC 443: \cite{chevalier1999} reviewed available
     observations and concluded that $n_a\sim 15$ cm$^{-3}$ and $v_s\sim 100$~\kms.
     We adopt $n_a=15$ cm$^{-3}$. \\
}
\end{deluxetable}
\end{document}